\documentclass[longauth]{aa}  

\usepackage{txfonts}
\usepackage{graphicx}
\usepackage{natbib}

\bibpunct{(}{)}{;}{a}{}{,} 

\def\hmpc{\,h^{-1}\rm{Mpc}}

\def\begf{\begin{figure}}
\def\endf{\end{figure}}

\def\mpcoh{\,h^{-1}{\rm Mpc}}

\begin{document} 

\title{The VIMOS Public Extragalactic Redshift Survey
  (VIPERS)\thanks{based on observations collected at the European
    Southern Observatory, Cerro Paranal, Chile, using the Very Large
    Telescope under programmes 182.A-0886 and partly 070.A-9007.  Also
    based on observations obtained with MegaPrime/MegaCam, a joint
    project of CFHT and CEA/DAPNIA, at the Canada-France-Hawaii
    Telescope (CFHT), which is operated by the National Research
    Council (NRC) of Canada, the Institut National des Sciences de
    l’Univers of the Centre National de la Recherche Scientifique
    (CNRS) of France, and the University of Hawaii. This work is based
    in part on data products produced at TERAPIX and the Canadian
    Astronomy Data Centre as part of the Canada-France-Hawaii
    Telescope Legacy Survey, a collaborative project of NRC and
    CNRS. The VIPERS web site is http://www.vipers.inaf.it/. }}

\subtitle{Full spectroscopic data and auxiliary
  information release (PDR-2)}

\author{
M.~Scodeggio\inst{\ref{iasf-mi}}       
\and L.~Guzzo\inst{\ref{brera},\ref{unimi}}      
\and B.~Garilli\inst{\ref{iasf-mi}}          
\and B.~R.~Granett\inst{\ref{brera},\ref{unimi}}          
\and M.~Bolzonella\inst{\ref{oabo}}      
\and S.~de la Torre\inst{\ref{lam}}       
\and U.~Abbas\inst{\ref{oa-to}}
\and C.~Adami\inst{\ref{lam}}
\and S.~Arnouts\inst{\ref{lam}} 
\and D.~Bottini\inst{\ref{iasf-mi}}
\and A.~Cappi\inst{\ref{oabo},\ref{nice}}
\and J.~Coupon\inst{\ref{geneva}} 
\and O.~Cucciati\inst{\ref{unibo},\ref{oabo}}           
\and I.~Davidzon\inst{\ref{lam},\ref{oabo}}   
\and P.~Franzetti\inst{\ref{iasf-mi}}   
\and A.~Fritz\inst{\ref{iasf-mi}}       
\and A.~Iovino\inst{\ref{brera}}
\and J.~Krywult\inst{\ref{kielce}}
\and V.~Le Brun\inst{\ref{lam}}
\and O.~Le F\`evre\inst{\ref{lam}}
\and D.~Maccagni\inst{\ref{iasf-mi}}
\and K.~Ma{\l}ek\inst{\ref{warsaw-nucl}}
\and A.~Marchetti\inst{\ref{iasf-mi}}
\and F.~Marulli\inst{\ref{unibo},\ref{infn-bo},\ref{oabo}} 
\and M.~Polletta\inst{\ref{iasf-mi},\ref{marseille-uni},\ref{toulouse}}
\and A.~Pollo\inst{\ref{warsaw-nucl},\ref{krakow}}
\and L.A.M.~Tasca\inst{\ref{lam}}
\and R.~Tojeiro\inst{\ref{st-andrews}}  
\and D.~Vergani\inst{\ref{iasf-bo}}
\and A.~Zanichelli\inst{\ref{ira-bo}}
\and J.~Bel\inst{\ref{cpt}}
\and E.~Branchini\inst{\ref{roma3},\ref{infn-roma3},\ref{oa-roma}}
\and G.~De Lucia\inst{\ref{oats}}
\and O.~Ilbert\inst{\ref{lam}}
\and H.~J.~McCracken\inst{\ref{iap}}
\and T.~Moutard\inst{\ref{halifax},\ref{lam}}  
\and J.~A.~Peacock\inst{\ref{roe}}
\and G.~Zamorani\inst{\ref{oabo}}
\and A.~Burden\inst{\ref{icg}}
\and M.~Fumana\inst{\ref{iasf-mi}}
\and E.~Jullo\inst{\ref{lam}}
\and C.~Marinoni\inst{\ref{cpt},\ref{inst-france}}
\and Y.~Mellier\inst{\ref{iap}}
\and L.~Moscardini\inst{\ref{unibo},\ref{infn-bo},\ref{oabo}}
\and W.~J.~Percival\inst{\ref{icg}}
}
  \offprints{Marco Scodeggio\\ \email{marcos@lambrate.inaf.it}}
\institute{
INAF - Istituto di Astrofisica Spaziale e Fisica Cosmica Milano, via Bassini 15, 20133 Milano, Italy \label{iasf-mi}
\and INAF - Osservatorio Astronomico di Brera, Via Brera 28, 20122 Milano
--  via E. Bianchi 46, 23807 Merate, Italy \label{brera}
\and  Universit\`{a} degli Studi di Milano, via G. Celoria 16, 20133 Milano, Italy \label{unimi}
\and INAF - Osservatorio Astronomico di Bologna, via Ranzani 1, I-40127, Bologna, Italy \label{oabo} 
\and  Aix Marseille Univ, CNRS, LAM, Laboratoire d’Astrophysique de Marseille, Marseille, France \label{lam}
\and INAF - Osservatorio Astrofisico di Torino, 10025 Pino Torinese, Italy \label{oa-to}
\and Laboratoire Lagrange, UMR7293, Universit\'e de Nice Sophia Antipolis, CNRS, Observatoire de la C\^ote d’Azur, 06300 Nice, France \label{nice}
\and Department of Astronomy, University of Geneva, ch. d’Ecogia 16, 1290 Versoix, Switzerland \label{geneva}
\and Dipartimento di Fisica e Astronomia - Alma Mater Studiorum Universit\`{a} di Bologna, viale Berti Pichat 6/2, I-40127 Bologna, Italy \label{unibo}
\and Institute of Physics, Jan Kochanowski University, ul. Swietokrzyska 15, 25-406 Kielce, Poland \label{kielce}
\and National Centre for Nuclear Research, ul. Hoza 69, 00-681 Warszawa, Poland \label{warsaw-nucl}
\and INFN, Sezione di Bologna, viale Berti Pichat 6/2, I-40127 Bologna, Italy \label{infn-bo}
\and Aix-Marseille Université, Jardin du Pharo, 58 bd Charles Livon, F-13284 Marseille cedex 7, France \label{marseille-uni}
\and IRAP,  9 av. du colonel Roche, BP 44346, F-31028 Toulouse cedex 4, France \label{toulouse} 
\and Astronomical Observatory of the Jagiellonian University, Orla 171, 30-001 Cracow, Poland \label{krakow} 
\and School of Physics and Astronomy, University of St Andrews, St Andrews KY16 9SS, UK \label{st-andrews}
\and INAF - Istituto di Astrofisica Spaziale e Fisica Cosmica Bologna, via Gobetti 101, I-40129 Bologna, Italy \label{iasf-bo}
\and INAF - Istituto di Radioastronomia, via Gobetti 101, I-40129, Bologna, Italy \label{ira-bo}
\and Aix Marseille Univ, Univ Toulon, CNRS, CPT, Marseille, France \label{cpt}
\and Dipartimento di Matematica e Fisica, Universit\`{a} degli Studi Roma Tre, via della Vasca Navale 84, 00146 Roma, Italy\label{roma3} 
\and INFN, Sezione di Roma Tre, via della Vasca Navale 84, I-00146 Roma, Italy \label{infn-roma3}
\and INAF - Osservatorio Astronomico di Roma, via Frascati 33, I-00040 Monte Porzio Catone (RM), Italy \label{oa-roma}
\and INAF - Osservatorio Astronomico di Trieste, via G. B. Tiepolo 11, 34143 Trieste, Italy \label{oats}
\and Institute d'Astrophysique de Paris, UMR7095 CNRS, Universit\'{e} Pierre et Marie Curie, 98 bis Boulevard Arago, 75014 Paris, France \label{iap} 
\and Department of Astronomy \& Physics, Saint Mary's University, 923 Robie Street, Halifax, Nova Scotia, B3H 3C3, Canada \label{halifax}
\and Institute for Astronomy, University of Edinburgh, Royal Observatory, Blackford Hill, Edinburgh EH9 3HJ, UK \label{roe}
\and Institute of Cosmology and Gravitation, Dennis Sciama Building, University of Portsmouth, Burnaby Road, Portsmouth, PO13FX \label{icg}
\and Institut Universitaire de France \label{inst-france}
}



 
  \abstract{We present the full public data release (PDR-2) of the VIMOS Public Extragalactic Redshift Survey (VIPERS), performed at the ESO VLT.  We release redshifts, spectra, CFHTLS magnitudes and ancillary information (as masks and weights) for a complete sample of 86,775 galaxies (plus 4,732 other objects, including stars and serendipitous galaxies); we also include their full photometrically-selected parent catalogue.  The sample is magnitude limited to $i_{\rm AB} \leq 22.5$, with an additional colour-colour pre-selection devised as to exclude galaxies at $z<0.5$. This practically doubles the effective sampling of the VIMOS spectrograph over the range $0.5<z<1.2$ (reaching 47\% on average), yielding a final median local galaxy density close to  $5\times 10^{-3}h^{3}{\rm Mpc}^{-3}$. The total area spanned by the final data set is $\simeq 23.5$~deg$^2$, corresponding to 288 VIMOS fields with marginal overlaps, split over two regions within the CFHTLS-Wide W1 and W4 equatorial fields (at R.A.$\simeq 2$ and $\simeq 22$ hours, respectively).
Spectra were observed at a resolution $R=220$, covering a wavelength range 5500-9500 \AA. Data reduction and redshift measurements were performed through a fully automated pipeline; all redshift determinations were then visually validated and assigned a quality flag. Measurements with a quality flag $\ge 2$ are shown to have a confidence level of 96\% or larger and make up 88\% of all measured galaxy redshifts (76,552 out of 86,775), constituting the VIPERS prime catalogue for statistical investigations.  
For this sample the {\sl rms} redshift error, estimated using repeated measurements of about 3,000 galaxies, is found to be $\sigma_z = 0.00054(1+z)$. All data are available at {\tt http://vipers.inaf.it} and on the ESO Archive.  
}
  

\maketitle
%

\section{Introduction}

Large photometric and spectroscopic galaxy surveys have played a key
role in building our current understanding of the Universe. At $z\le
0.2$, the 2dFGRS \citep{colless03} and SDSS \citep{york00,sdss_dr7} redshift surveys have assembled samples of over a million objects, precisely characterising large-scale structure and galaxy properties in the nearby Universe on scales ranging from 0.1 to $100\mpcoh$.  The SDSS has then extended its reach, first by using luminous red galaxies (LRG) to push to $z\simeq 0.35$ \citep[SDSS-II:][]{eisenstein11,sdss_dr9}, and more recently (to $z\simeq 0.5$) by using a more heterogeneous set of colour-selected objects to trace large volumes of the Universe in a highly effective way, notwithstanding a rather dilute sampling of the total galaxy population \citep[SDSS-III BOSS:][]{alam15}. [A more complete account of the development of galaxy redshift surveys over the past two decades was given in \citealt{guzzo14}].  

The VIMOS Public Extragalactic Redshift Survey (VIPERS) adopted the original broad approach of SDSS-I, transposed to the redshift range $0.5 < z < 1.2$, essentially extending to a much larger volume of the Universe the exploration initiated with smaller-area  VIMOS precursors, i.e. VVDS \citep{lefevre13,garilli08} and zCOSMOS \citep{lilly09}. In practice, VIPERS was conceived to obtain a large-volume, dense sample of the general galaxy population, characterised by a simple,  broad selection function, complete to a given flux limit within a well-defined redshift range and complemented by extended photometric information. 

This paper accompanies the Public Data Release 2 (PDR-2) of the complete VIPERS data set and is organised as follows: in Sect.~\ref{sec:design} we summarise the survey design and scope, which we discussed in detail in the papers by \citet{guzzo14} and \citet{garilli14}, which accompanied the first data release (PDR-1); in Sect.~\ref{sec:sky} we present the final survey mask and completeness estimates, while redshift measurements are summarised in Sect.~\ref{sec:redshift_measure}, and the overall properties of the PDR-2 sample are presented in Sect.~\ref{sec:dataset}.


\section{Summary of survey design and execution}

\subsection{Survey design}
\label{sec:design}

VIPERS was designed to sample, at a median redshift $z \simeq 0.7$, a volume 
comparable to the one covered by redshift surveys mapping the local Universe 
(2dFGRS and SDSS), with a similarly high sampling density of the galaxy population. 
To achieve useful spectral quality in a limited exposure time using the VIMOS spectrograph \citep{lefevre03}, a relatively bright limit of $i_{\rm AB} \leq 22.5$ was adopted, and this generated two main issues for efficient sample selection. At this depth, many galaxies will lie below the redshift range of interest, i.e. $0.5<z<1.2$ \citep[see for example][]{lefevre05, lefevre13, lilly09}; also, it was known from previous similar studies such as the VVDS-Wide \citep{garilli08} that such a purely magnitude-limited sample would suffer from approximately 30\% stellar contamination. Here we give a brief summary of the steps taken to overcome these difficulties (see \citealt{guzzo14} and \citealt{garilli14} for fuller details).

The VIPERS target selection was derived from the `T0005' release of the CFHTLS Wide photometric survey, completed and improved using the subsequent T0006 release. A preliminary multi-band catalogue, including all objects with extinction-corrected apparent magnitude $i_{\rm AB} \leq 22.5$, was built starting from the individual CFHTLS 1-deg$^2$ tiles.  
Particular care was taken to verify the homogeneity of these original single-tile catalogues: by analyzing the colour-colour stellar locus within each such catalogue, we were able to identify significant tile-to-tile offsets in the photometric zero-points for different photometric bands. To ensure that the final VIPERS parent photometric catalogue was as spatially homogeneous as possible, a tile-to-tile offset correction to the observed colours was estimated and applied. As discussed extensively by \citet{guzzo14}, this offset was obtained by comparing the position in colour space of the (well-defined) stellar locus with that of a reference tile (the one overlapping the VVDS F02 survey field, \citealt{lefevre13}).

This homogenisation of galaxy colours over the full area was particularly crucial for the subsequent removal of low-redshift galaxies (nominally $z<0.5$), which was implemented via a robust colour-colour selection in the $(r-i)$ vs $(u-g)$ plane, tuned using the VVDS complete redshift data.  Details can be found in \citet{guzzo14}. 
%
%
Finally, stellar objects were removed using a combination of two methods: for objects brighter than $i_{\rm AB}=21.0$, stars were identified on the basis of their half-flux radius, as measured on the $i$-band CFHTLS images; for fainter objects, a combination of image size and SED fitting of the 5-band CFHTLS photometry was used (see Appendix A of \citealt{guzzo14}, and Section 2.1 of \citealt{garilli14}). 

Overall, some 21\% of the objects in the total photometric catalogue have been removed because they were classified as stars, 32\% were removed because they were classified as low redshift galaxies, and the remaining 47\% became the VIPERS parent photometric sample, which was then supplemented with a small additional sample of AGN candidates, chosen from objects that were initially classified as stars on the basis of a colour-colour criterion (see Section 2.2 of \citealt{garilli14}). This sample contributes on average 2-5 objects per VIMOS quadrant (against about 90 galaxy targets) with negligible impact on the galaxy selection function.  In the PDR-2 catalogue these additional objects can be easily identified and separated from the main galaxy sample through an appropriate keyword, as described in Sect.~\ref{sec:dataset}.

   \begin{figure}
   \centering
    \includegraphics[width=8cm]{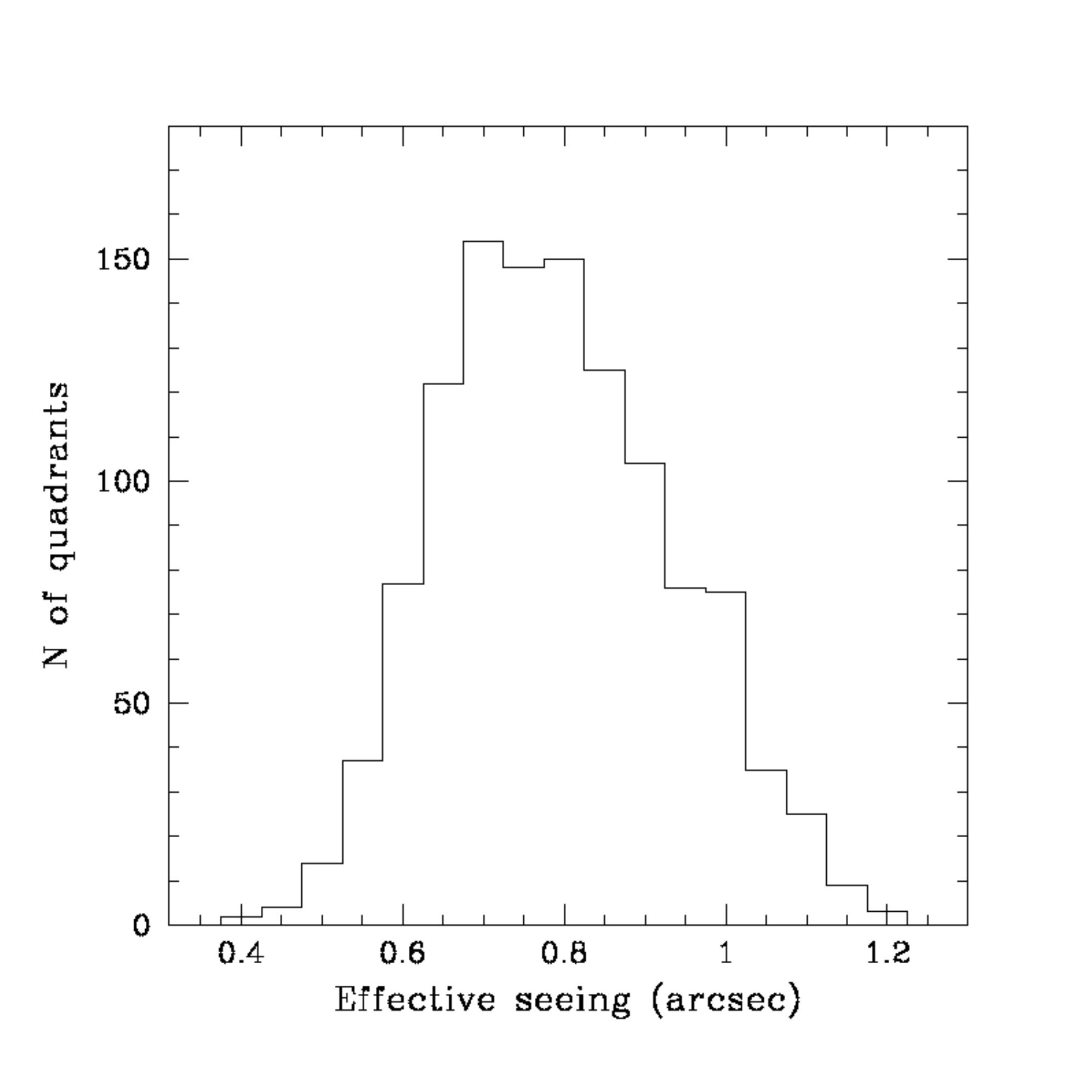}
     \caption{The effective seeing distribution for the VIPERS observations. 
     The seeing value is obtained from the measurement of the FWHM size of the
     spectral traces for bright objects in the spectroscopic exposures.
              }
         \label{fig:seeing_distribution}
   \end{figure}

\subsection{Improvements in CFHTLS photometry during the construction of VIPERS}
\label{sec:cfhtls_t07}

The tile-to-tile colour shifts in the T0005 data discussed in the previous section were a clear evidence that the initial global photometric calibration could be significantly improved. Such a step forward was provided by the CFHTLS T0007 revision \citep{hudelot12}. For VIPERS, the most important feature of T0007 compared to previous releases is that each tile in the CFHTLS was rescaled to an absolute calibration provided by a new dedicated survey of calibrators carried out at the CFHT. In addition, in order to ensure that seeing variations between tiles and filters were correctly accounted for, aperture fluxes were rescaled to allow for the seeing of each individual tile. These aperture fluxes have then become the basis of the new photometric catalogue for VIPERS, since it has been shown that Kron (mag auto) flux estimates provide less accurate colour estimates, which lead, among other things, to worse photo-$z$'s \citep{moutard16a, hildebrandt12}. 
A new photometric catalogue was therefore created, based on Terapix T0007 isophotal aperture magnitudes, with the addition of UV photometry from GALEX \citep{Martin2005}, and NIR $K_s$-band photometry from WIRCam \citep{Puget2004} or from VISTA \citep{Emerson2004}, obtained as part of the VIPERS Multi-Lambda Survey (VIPERS-MLS, \citealt{moutard16a}) or the VISTA Deep Extragalactic Observations (VIDEO, \citealt{Jarvis2013}), respectively. The isophotal magnitudes were then corrected to pseudo-total ones using an aperture correction for each individual object, obtained as the average of the aperture corrections obtained for the $g,r,i$, and $K_s$ bands. Details can be found in \cite{moutard16a}.

   \begin{figure*}
   \centering
   \includegraphics[width=\hsize]{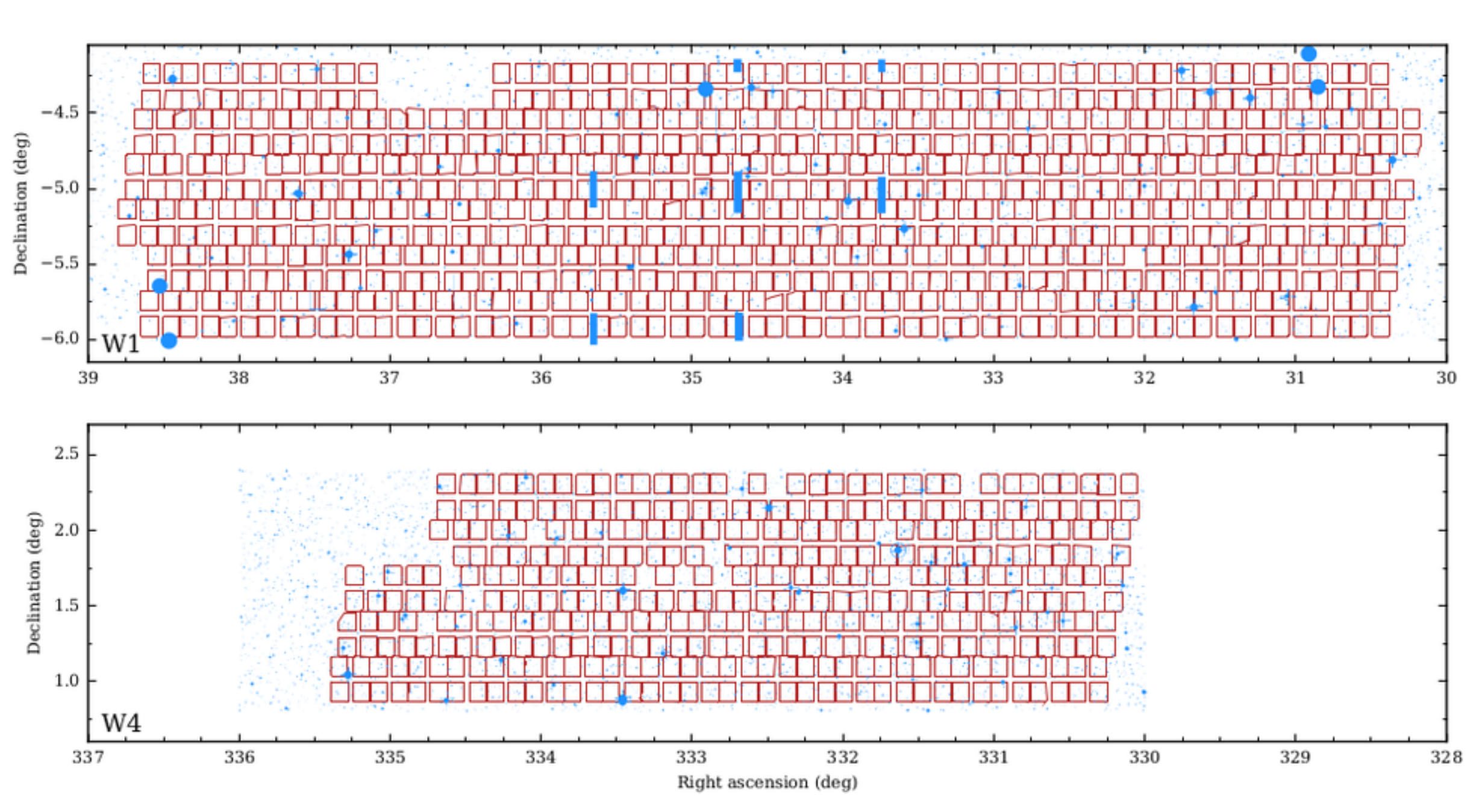}
     \caption{The layout on the sky of all pointings that  contribute to the PDR-2 final release, for the two fields W1 and W4, superimposed on the photometric survey mask. The contours of each of the four quadrants that comprise all VIMOS pointings are shown in red. The blue (grey) areas in the background correspond to areas where the parent photometry is corrupted or observations are not possible due to the presence of foreground objects, particularly bright stars and their diffraction spikes. Note that in this plot failed quadrants and other features introduced by the `spectroscopic mask' are not indicated (see Section~\ref{sec:spectro_mask} and Figs~\ref{fig:photo_mask} and \ref{fig:tsr}).
              }
         \label{fig:photo_mask_big}
   \end{figure*}

This new photometric catalogue became available after VIPERS was well under way, and it was therefore decided not to replace the original parent VIPERS catalogue, because such a substitution would have turned the original well-defined magnitude limit into a somewhat fuzzy limit, resulting from the approximately 0.05 mags scatter in the magnitude comparison between the T0005 and T0007 photometry (at the 22.5 magnitude limit of the catalogue). A choice was therefore made to match the two catalogues, and to provide T0007 photometry for all the objects in the original T0005-based catalogue.
The matching was carried out using a match circle of 0.6 arcsec, which ensures a 97\% matching success rate, with an almost null incidence of multiple object matches. Although the original VIPERS catalogue is limited to $i_{\rm AB} \leq 22.5$, the matching was carried out limiting both the old and the new catalogue to $i_{\rm AB} \leq 23.0$, to avoid the scatter at the catalogue limit to affect the resulting match. For the small fraction of objects that could not be satisfactorily matched (mostly due to small differences in the source de-blending procedure), we computed pseudo-T0007 magnitudes: we estimated the median offset between the T0005 and T0007 catalogues for each photometric band and each CFHTLS parent tile using objects with $i_{\rm AB} \leq 21.0$, and added this offset to the original T0005 magnitudes (the magnitude uncertainty was kept equal to the T0005 value).   

Areas in the photometric catalogue with poor quality, corrupted source extraction, or bright stars are described by a binary mask, as discussed in Sect.~\ref{sec:sky}.

\subsection{Spectroscopic observations}
\label{sec:obs}

All VIPERS observations were carried out using VIMOS (ESO VIsible Multi-Object 
Spectrograph), on `Melipal', Unit 3 of the ESO Very Large Telescope (VLT) -- see \citet{lefevre03}. VIMOS is a 4-channel imaging spectrograph; each channel (a `quadrant') covers $ \sim 7 \times 8 $ arcmin$^2$ for a total field of view (a `pointing') of $\sim 224$ arcmin$^2$. Each channel is a complete spectrograph with the possibility of inserting $30\times30$ cm$^2$ slit masks at the entrance focal plane, as well 
as broad-band filters or grisms. The precise sizes of the quadrants are in principle all slightly different from each other: the four channels of VIMOS all differ slightly, and they also changed with time during the survey development, when the VIMOS CCDs were refurbished (see below). There is also variation from one pointing to another, e.g. due to vignetting by the guide star probe. All these pieces of information are quantified accurately by the mask files associated with the PDR-2 release, which we discuss in the following section.  

The pixel scale on the CCD detectors is 0.205 arcsec/pixel, providing excellent sampling of the Paranal mean image quality and Nyquist sampling for a slit 0.5 arcsec in width. For VIPERS, we used a slit width of 1 arcsec, together with the `low-resolution red' (LR-Red) grism, resulting in a spectral resolution $R\simeq 220$ at the centre of the wavelength range covered by this grism (i.e. $\sim 5500-9500$ \AA). 
In summer 2010, VIMOS was upgraded with new red-sensitive CCDs in each of the four channels, as well as with a new active flexure compensation system. The reliability of the mask exchange system was also improved \citep{hammersley10}. The original thinned E2V detectors were replaced by twice-thicker E2V devices, considerably lowering the fringing and increasing the global instrument efficiency by up to a factor 2.5 (one magnitude) in the redder part of the wavelength range.  This upgrade significantly improved the average quality of VIPERS spectra, resulting in a significantly higher redshift measurement success rate.  

The complete VIPERS survey consists of 288 VIMOS pointings, 192 over the W1
area, and 96 over the W4 area of the CFHTLS, overlapping a total sky area of about 23.5 square degrees. Due to the specific footprint of VIMOS, failed quadrants and masked regions, this corresponds to an effectively covered area of 16.3 square degrees. The number of slits in the spectroscopic masks ranged from 60 to 121 per VIMOS quadrant, with a median value of 87, for a total of 96,929 slits over the whole survey. For nine pointings, observations were repeated because the original observation was carried out under sub-optimal seeing and/or atmospheric conditions, while four pointings already observed before the VIMOS 2010 upgrade were re-observed as part of the related re-commissioning. Overall, observations were carried out starting in November 2008, and were completed by December 2014. For 23 pointings (6 in the W1 area, and 17 in the W4 one) some mask insertion problem prevented the acquisition of useful spectroscopic data in one of the four VIMOS quadrants, leaving some small ``holes'' in the survey sky coverage (see Fig.~\ref{fig:photo_mask_big}). These are termed "failed" quadrants. Airmass during the observations ranged from 1.06 to 1.44, with  a median value of 1.14, while the effective seeing (measured directly from the observed size of the reference objects used to align the VIMOS masks) ranged from 0.41 to 1.21 arcsec, with a median value of 0.78 arcsec. Fig.~\ref{fig:seeing_distribution} shows the distribution of these effective seeing values.

   \begin{figure}
   \centering
    \includegraphics[width=8cm]{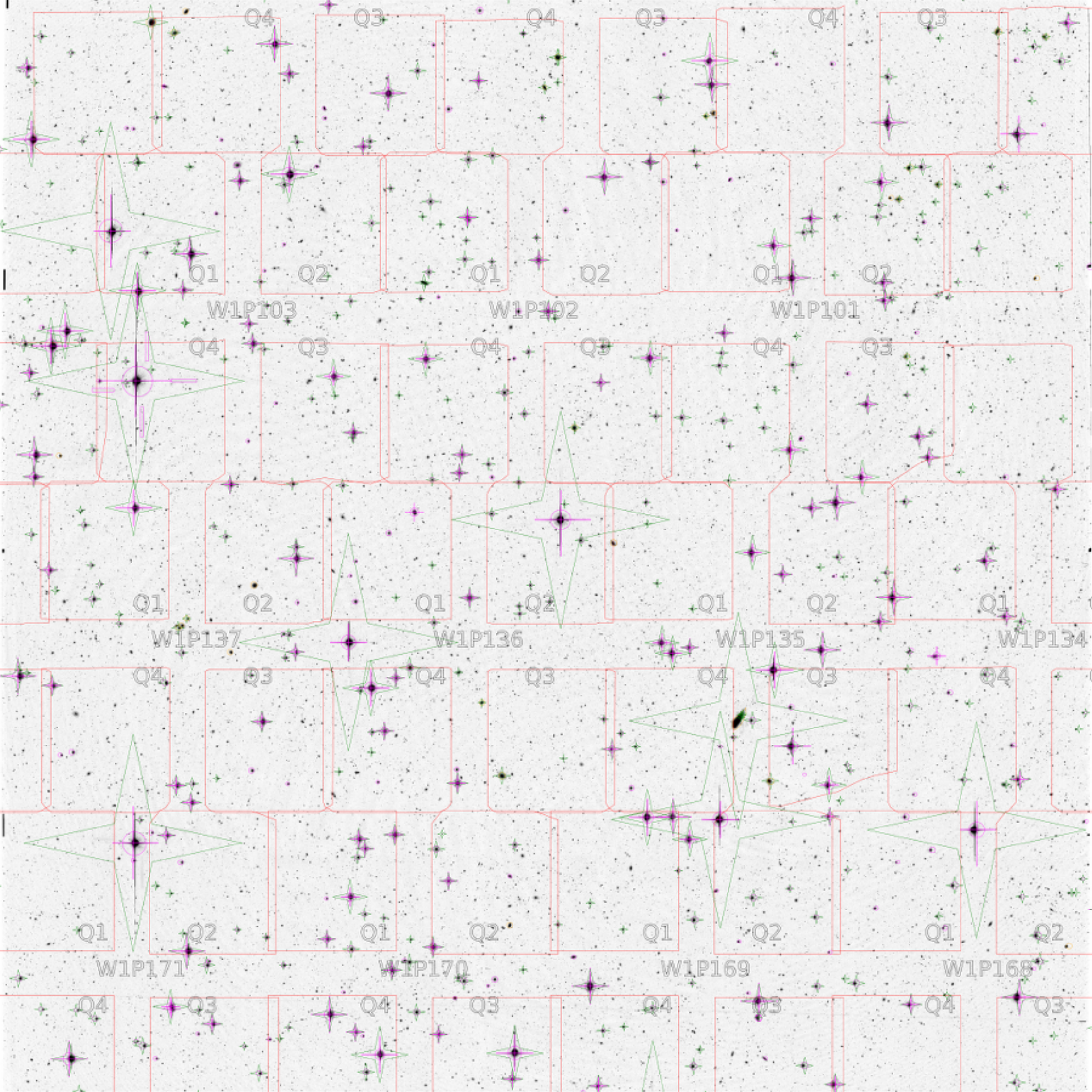}
     \caption{A 1 deg$^2$ detail of the masks developed for VIPERS.  The revised photometric mask built for VIPERS corresponds to the magenta circles and cross patterns; for comparison, the original, more conservative mask distributed by Terapix is shown in green. The quadrants that make up the 
        VIPERS pointings are plotted in red. In the background is the
        CFHTLS T0006 $\chi^2$ image of the field 020631$-$050800
        produced by Terapix.  Note the significant gain in usable sky
        obtained with the new VIPERS-specific mask.
              }
         \label{fig:photo_mask}
   \end{figure}


\section{Sky coverage, angular selection functions and completeness}
\label{sec:sky}

The VIPERS angular selection function is the result of the combination of several different angular completeness functions. Two of these are binary masks (i.e. describing areas that are fully used or fully lost). The first mask (that we call the photometric mask) is related to defects in the parent photometric sample (mostly areas masked by bright stars) and the other (that we call the spectroscopic mask) to the specific footprint of VIMOS and how the different pointings are tailored together to mosaic the VIPERS area. The other completeness functions are provided instead on a per-galaxy basis:
1) within each of the four VIMOS quadrants on average only 47\% of the available targets satisfying the selection criteria are actually placed behind a slit and observed, defining what we call the Target Sampling Rate; 2) since the set of available targets is defined based on the observed colour, as discussed in Sect.~\ref{sec:design}, a Colour Sampling Rate is needed to keep this selection effect into consideration; 
3) varying observing conditions and technical issues determine a variation from quadrant to quadrant of the actual number of redshifts measured with respect to the number of targeted galaxies, while our capability to measure the redshift depends on intrinsic galaxy parameters, as we shall discuss in Sect.~\ref{sec:ssr} when introducing the Spectroscopic Success Rate.

Detailed knowledge of all these contributions is a crucial ingredient for computing any quantitative statistics of the galaxy distribution, as e.g. its first and second moments (i.e. luminosity / stellar mass functions and two-point correlation functions, respectively). 

   \begin{figure}
   \centering
   \includegraphics[width=8cm]{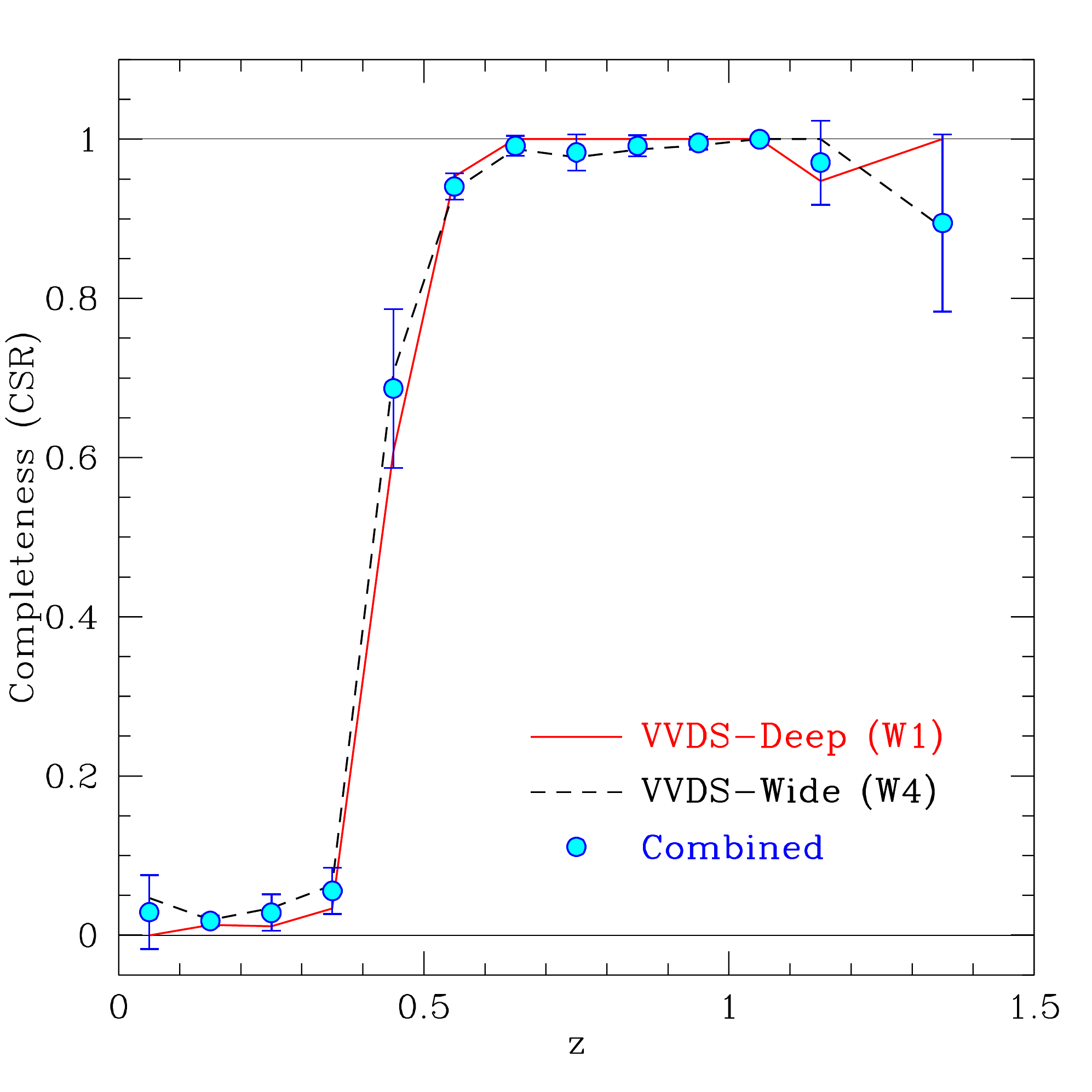}
     \caption{Estimate of the Colour Sampling Rate (CSR) of VIPERS 
     The plot shows the fraction of galaxies that are selected by the VIPERS colour-colour criteria as a function of redshift, when applied to a sample of galaxies from the VVDS-Deep and VVDS-Wide surveys \citep{lefevre13}. This is a highly significant test, given that the original colour-colour boundaries to select VIPERS targets were calibrated on the same VVDS data.  Both W1 and W4 fields provide consistent selection functions, yielding a colour selection function that is essentially unity above $z=0.6$ and can be consistently modelled in the transition region $0.4<z<0.6$.
              }
         \label{fig:CSR}
   \end{figure}


   \begin{figure*}
   \centering
   \includegraphics[width=\textwidth]{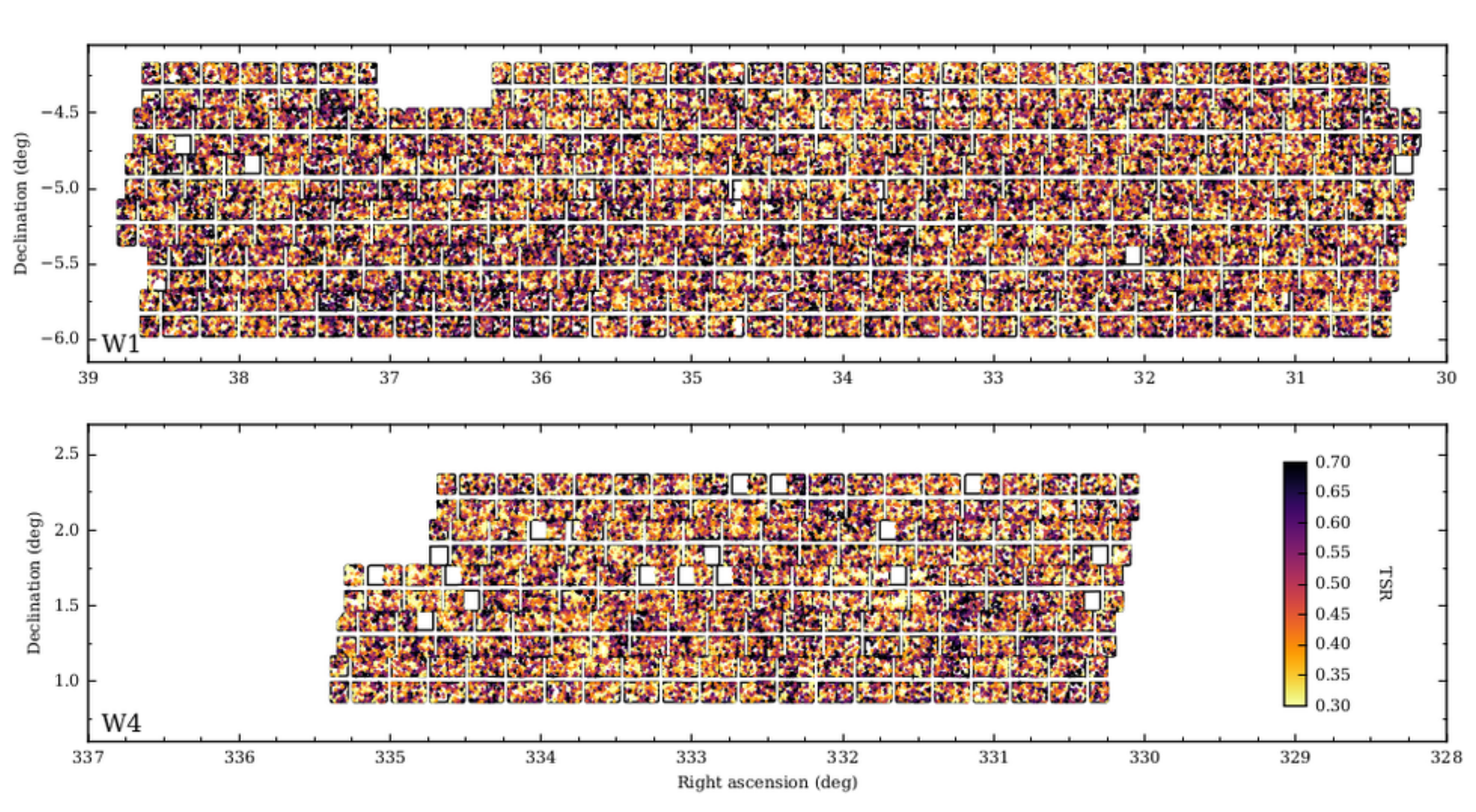}
     \caption{The angular distribution over the survey areas of the Target Sampling Rate (TSR, colour scale), estimated locally for each galaxy as described in the text. In this plot, each value has been smoothed on a scale of 2 arcmin to enhance the (inverse) relationship of the TSR with the  projected large-scale structure in the galaxy distribution.
              }
         \label{fig:tsr}
   \end{figure*}

\subsection{CFHTLS-VIPERS photometric mask}
\label{sec:photo_mask}

The photometric quality across the CFHTLS images is tracked with a set of masks that account for imaging artefacts and non-uniform coverage. We use the masks to exclude regions from the survey area with corrupted source extraction or degraded photometric quality. The masks consist primarily of patches around bright stars 
($B_{\rm Vega}<17.5$) owing to the broad diffraction pattern and internal reflections in the telescope optics. At the core of a saturated stellar halo there are no reliable detections, leaving a hole in the source catalogue, while in the halo and diffraction spikes spurious sources may appear in the catalogue due to false detections. We also add to the mask extended extragalactic sources that may be
fragmented into multiple detections or that may obscure potential VIPERS sources. The details of the revised photometric mask construction were given in \citet{guzzo14} and a visual rendition of the two W1 and W4 masks is given in Fig.~\ref{fig:photo_mask_big}, while Fig.~\ref{fig:photo_mask} provides a zoom into a smaller area, in particular showing the details of the custom-developed VIPERS photometric mask, compared to the original CFHTLS mask.

It is important to stress here that a small fraction of spectroscopically observed galaxies actually fall within regions forbidden by the photometric mask. These are objects for which, typically, one of the photometric bands had too large an error to be acceptable, but were nevertheless observed as fillers. As such, in any computation of spatial statistics the photometric and spectroscopic masks must be applied not only to any auxiliary random sample (as typically needed for two-point clustering measurements), but also re-applied to the observed spectroscopic catalogue itself. This is required in order to trim these little `leakages' within a few specific areas.

\subsection{VIPERS spectroscopic masks}
\label{sec:spectro_mask}

The general layout of VIMOS is well known, but the precise geometry of each quadrant has to be specified carefully for each observation, in order to perform precise clustering measurements with the VIPERS data.  For example, although it rarely happens, a quadrant may be partly vignetted by the VLT guide probe arm; in addition, the size and geometry of each quadrant changed slightly between the pre- and post-refurbishment data (i.e. from mid-2010 on), due to the dismounting of the instrument and the technical features of the new CCDs.  We therefore had to build our own extra mask for the spectroscopic data, accounting for all these aspects at any given point on the sky covered by the survey.

The masks for the W1 and W4 data were constructed from the pre-imaging observations by running an image analysis routine that identifies `good' regions within those images. 
First, a polygon is defined that traces the edge of the image. The mean and variance of the pixels are computed in small patches at the vertices of the polygon, and these measurements are compared to the statistics at the centre of the image. The vertices of the polygon are then iteratively moved inward toward the centre until the statistics along the boundary are within an acceptable range of those measured at the centre.  The boundary that results from this algorithm is used as the basis for the field geometry. The polygon is next simplified to reduce the vertex count: short segments that are nearly co-linear are replaced by long segments. The World Coordinate System information in the fits header is used to convert from pixel coordinates to sky coordinates.  Each mask was then examined by eye. Features due to stars at the edge of an image were removed, wiggly segments were straightened and artefacts due to moon reflections were corrected. The red lines in Fig.~\ref{fig:photo_mask} show the detailed borders of the VIMOS quadrants, describing the spectroscopic mask.

\begin{figure*}
\begin{center}
\includegraphics[width=\hsize]{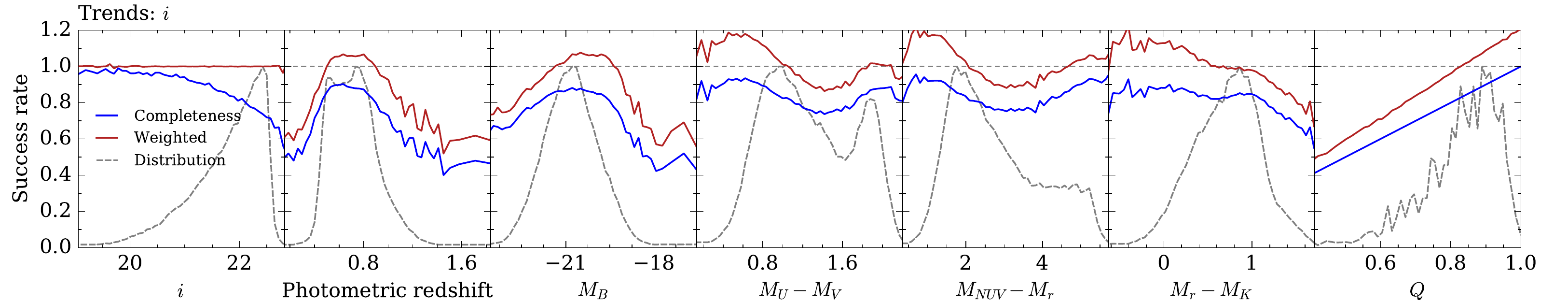}
\includegraphics[width=\hsize]{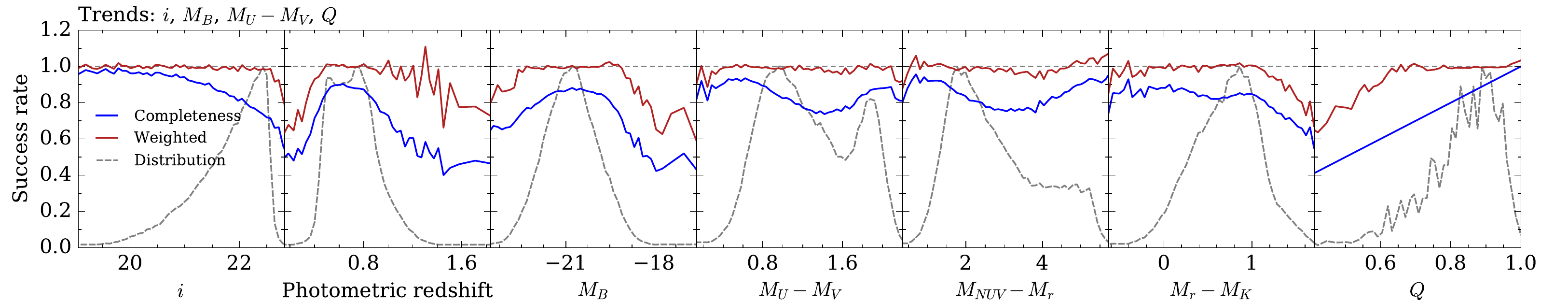}
\end{center}
\caption{The spectroscopic success rate (SSR) as a function of different observed photometric properties (blue solid curve), compared to the result of applying the weight to correct incompleteness (red solid curve). The two rows of plots show how the completeness correction changes when one only considers a simple SSR dependence on the selection $i$-band magnitude (top), or rather includes the more subtle dependencies on observed colours and, in particular, quality of the specific VIMOS quadrant (`Q' parameter, quantified via the mean SSR for all galaxies in that quadrant).  The differential distribution of each parameter is also plotted in each panel (dashed curve).
}
  \label{fig:ssr_param}
\end{figure*}

\subsection{Colour Sampling Rate}
\label{sec:csr}

The completeness of the colour-colour pre-selection applied to ideally isolate $z>0.5$ galaxies using the CFHTLS corrected photometry, has been quantified and discussed in \citet{guzzo14}. Using the data from the VVDS survey, the Colour Sampling Rate (CSR) was estimated as a function of redshift. This is shown in Fig.~\ref{fig:CSR} [originally from  \citet{guzzo14}], which we reproduce here for completeness: from this figure it is quite clear how the VIPERS catalogue is virtually 100\% complete above $z = 0.6$, when compared to a corresponding purely magnitude-limited sample.  

\begin{figure*}
\begin{center}
\includegraphics[width=18cm]{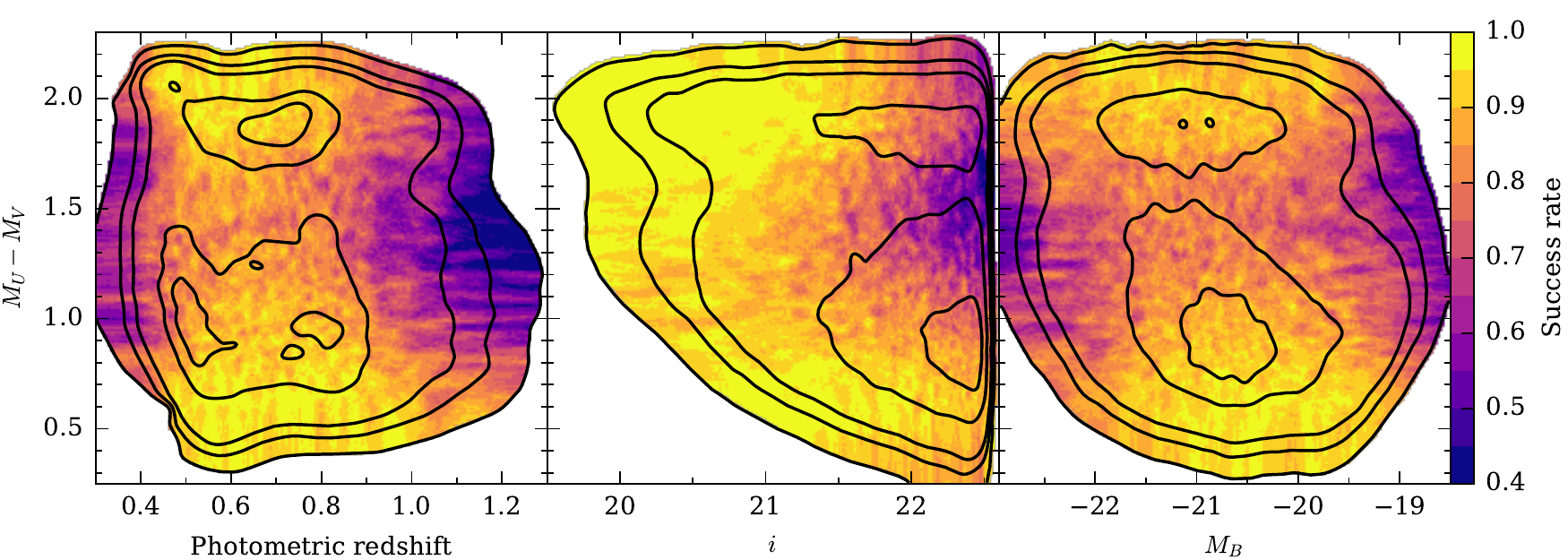}
\end{center}
\caption{2D plots of the SSR in the photometric parameter space. 
We show the dependence on the rest-frame $U-V$ colour jointly with three photometric parameters: photometric redshift (left), apparent $i$-band magnitude (centre), and rest-frame $B$-band magnitude (right).  The solid contours contain 10, 50, 90 and 99\% of the sample.
  \label{fig:ssr}}
\end{figure*}

\subsection{Target Sampling Rate}
\label{sec:tsr}

A Multi-Object Spectrograph (MOS) survey inevitably has to deal with the limitation of MOS slits creating a shadow effect in the targeting of potential sources that is strongly density-dependent. In practice the high-density peaks of the projected galaxy density field are under-sampled with respect to the low-density regions, because the MOS slit length imposes a minimum angular pair separation in the spectroscopic target selection. In VIPERS, this was performed using the SPOC algorithm \citep[][within the VMMPS software distributed by ESO]{bottini05}, which maximises the number of slits observed in each quadrant. As a result, (a) very close pairs below a certain scale are practically unobservable; (b) the angular distribution of slits is more uniform than the underlying galaxy distribution.
In VIPERS, the first effect suppresses angular clustering below a scale of 5 arcsec, producing a scale-dependent damping of the observed clustering below $\simeq 1 \hmpc$; the second is instead responsible for a nearly scale-independent reduction of the two-point correlation function amplitude above this scale. These effects and their correction are discussed in detail in the parallel paper by \citet{pezzotta16}. The method builds upon the original approach of \citep{delatorre13}, by up-weighting galaxies on the basis of the Target Sampling Rate (TSR), namely
\begin{equation}
w_i=\frac{1}{\mathrm{TSR}_i}. 
\label{eq:tsr_weight}
\end{equation}
In \citet{delatorre13}, however, the TSR was evaluated only on a quadrant to quadrant basis so that all the targeted galaxies falling in the same quadrant are up-weighted by the same factor. This procedure does not recover the total missing power, since it does not account for the above effects on sub-quadrant scales. The new corrective approach is similar, but uses a local TSR that accounts much more effectively for the angular inhomogeneity of the  selection function. This is estimated for each galaxy as the ratio of the local surface densities of target and parent galaxies (i.e. before and after applying the target selection), properly estimated and then averaged within an aperture of appropriate shape and size. If we call these quantities $\delta_i^p$ and $\delta_i^s$, then the TSR$_i$ is defined as
\begin{equation}
\mathrm{TSR}_i=\frac{\delta_i^s}{\delta_i^p}.
\label{eq:tsr_formula}
\end{equation}
A continuous $\delta$ field is obtained, starting from the discrete surface distribution, by first using a classical Delaunay tessellation to get the density at the position of each galaxy, and then linearly interpolating. This is finally integrated around the position of each observed galaxy 
within a rectangular aperture with size $60\times 100\,\mathrm{arcsec^2}$ to obtain the local values of $\delta_i^s$ and $\delta_i^p$. It can be shown \citep{pezzotta16} that a rectangular aperture more efficiently accounts for the angular anisotropy in the distribution of targets within a quadrant introduced by the shadowing effect of the MOS slits. 

The resulting distribution of the TSR values over the survey regions is shown in Fig.~\ref{fig:tsr}. Thanks to the adopted strategy (i.e. having discarded through the colour selection almost half of the magnitude-limited sample lying at $z<0.5$), the average TSR of VIPERS is $\simeq 47\%$, a high value that represents one of the specific important features of VIPERS. For a comparison, with the VVDS-Wide sample, selected to the same magnitude limit, but without colour pre-selection and star rejection, the sampling rate was about 23\% \citep{garilli08}, i.e. half of what we have achieved here. We remark how the TSR essentially mirrors the intrinsic fluctuations in the number density of galaxies as a function of position on the sky, and how single quadrants sometimes have a strong internal inhomogeneity in the sampling of galaxies.

\subsection{Spectroscopic Success Rate}
\label{sec:ssr}

We quantify the VIPERS redshift measurement success via the Spectroscopic Success Rate (SSR), which is defined as the ratio between the number of objects for which we have successfully measured a redshift $N_{\rm success}$ and the number of objects targeted by the spectroscopic observations $N_{\rm target}$. We define the success of a redshift measurement on the basis of the redshift quality flag discussed in detail in Sect.~\ref{sec:redshift_flags}; generally we adopt the 95\% measurement confidence threshold (see Sect.~\ref{sec:redshift_reliability}) and accept flags 2, 3, 4 and 9 as  markers of a successful measurement. Targeted objects for which the spectral extraction completely failed (often these are spurious objects in the photometric catalogue) are not counted among the targets. Additionally, we correct for the stellar contamination by subtracting the number of spectroscopically confirmed stars $N_{\rm star}$ from the numerator and denominator.  The SSR estimator is therefore:
\begin{equation}
{\rm SSR} = \frac{N_{\rm success} - N_{\rm star}}{N_{\rm target} - N_{\rm star}}.
\end{equation}

   \begin{figure*}
   \centering
   \includegraphics[width=17cm]{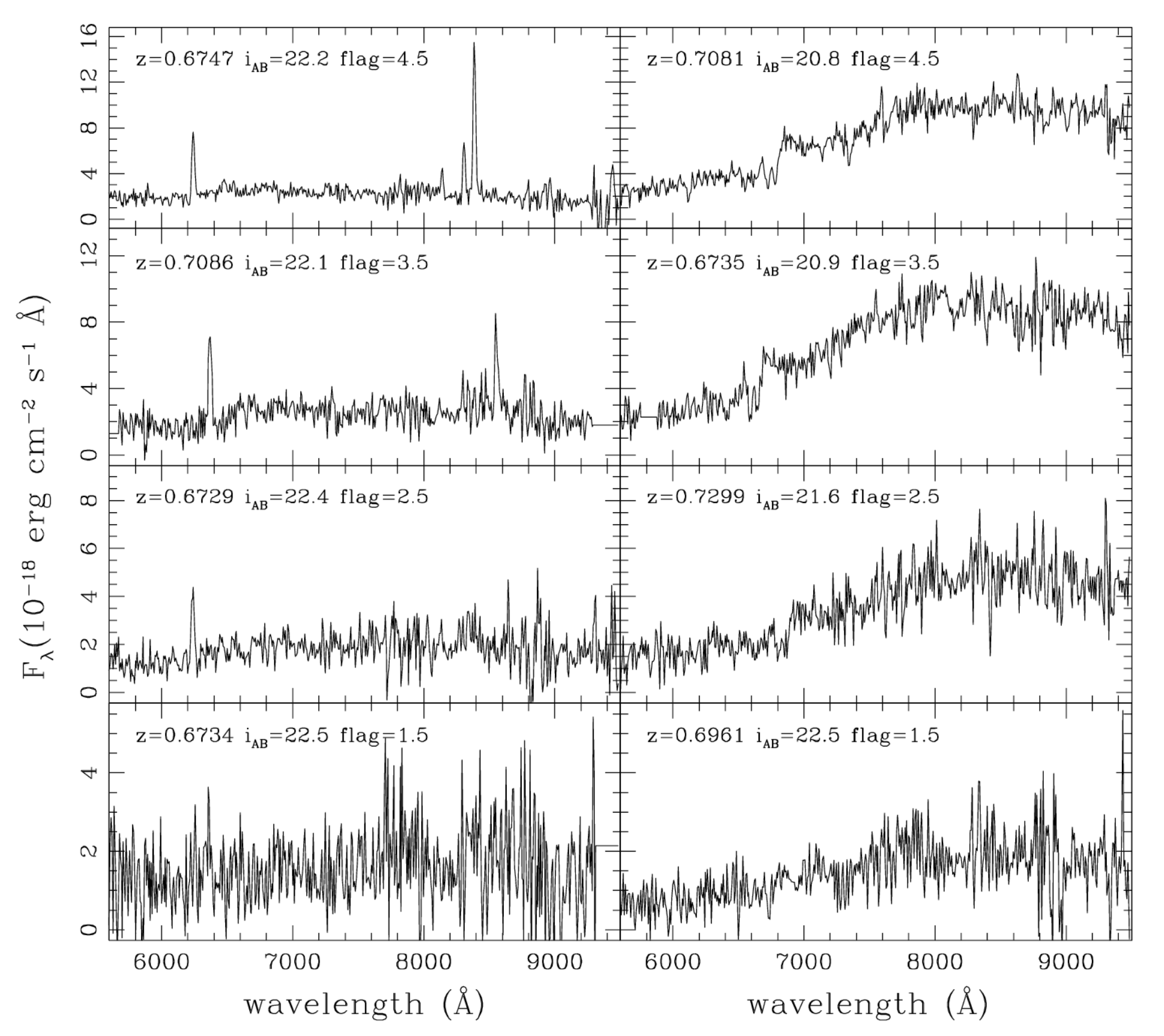}

      \caption{Examples of VIPERS spectra: one late-type and one early-type galaxy spectrum
      is shown for the different redshift measurement quality flags. The measured redshift
      for all objects is close to $z=0.7$, the peak of the VIPERS redshift distribution
              }
         \label{fig:spectra}
   \end{figure*}

The ability to  measure a redshift with confidence depends on a number of factors, starting from the observing conditions for a given survey pointing, and the apparent flux of a given galaxy, but also including intrinsic galaxy properties, such as its spectral type and redshift. The top part of Fig.~\ref{fig:ssr_param} shows that, if we use just the galaxy apparent $i$-band magnitude to parametrize the SSR, we cannot successfully reproduce the SSR dependence on other parameters, such as the galaxy rest-frame colour. When instead we use multiple parameters, including the apparent magnitude, the rest-frame colour, the galaxy $B$-band luminosity, and the overall quality of the specific VIMOS quadrant (quantified via the mean SSR for all galaxies in that quadrant), we obtain a significant improvement on the SSR capability of describing the VIPERS sample, as shown in the bottom part of Fig.~\ref{fig:ssr_param}. In both rows of plots in that figure we show the SSR as a function of different observed photometric properties (blue solid curve), compared to the result of applying the SSR weight to correct for incompleteness (red solid curve) the total VIPERS sample. Rest-frame properties (galaxy luminosity and colour) in this case are computed on the basis of the galaxy photometric redshift, in order to enable their computation also for galaxies without a spectroscopic redshift measurement (that enter into the denominator of eq. 3). Photometric redshift estimates are taken from \cite{moutard16a}, and have a typical accuracy of $\sigma_z \leq 0.04$, with a fraction of catastrophic failures smaller than 2\%. 

The SSR is computed adaptively using a nearest-neighbour algorithm. Depending on which parameters we want to use to parametrize the SSR, we build an $N$-dimensional dataset (with $N=1$ when we use only the apparent magnitude, and $N=4$ when we add also the rest-frame colour, the luminosity, and the quadrant quality);  then, for each object in this $N$-dimensional space, we determine the distance $R_K$ to its $K^{\rm th}$ nearest neighbour (we use $K=100$). We then count the number of sources in the successfully measured sample that are contained within the radius $R_K$: $N_{\rm success}(\le R_K)$. The SSR at the specified point is given by the fraction $SSR = N_{\rm success}(\le R_K) / K$.  Distances in this $N$-dimensional space are computed using the rank distance measure (i.e. the ranks of each parameter in the sample are used when computing separations).  

Fig.~\ref{fig:ssr} shows the bivariate distribution of SSR values as a function of  rest-frame colour and redshift, apparent $i$-band magnitude, and rest-frame $B$-band luminosity, respectively. The mean SSR is about 83\%, but it is clear from this figure that complex SSR variations exist as a function of galaxy properties. An obvious apparent magnitude trend is clearly visible in the figure's middle panel, but all panels quite clearly show how the lowest SSR values are characteristic of galaxies with intermediate rest-frame colour. These are objects whose spectra contain neither strong emission lines (as would be seen in the bluest part of the sample) nor a strong \smash{4000 \AA} break (as would be seen in the reddest part of the sample). This general feature was already observed for the zCOMOS bright survey \citep[][see their Fig.2]{lilly09}.

\section{Redshift measurements, confidence flags and statistical uncertainties}
\label{sec:redshift_measure}

Fig.~\ref{fig:spectra} shows a representative set of spectra of different quality, as available at the end of the automatic data reduction pipeline of VIPERS. This, together with the procedure for redshift validation, have been extensively described in \citet{garilli10}, \citet{garilli12}, \citet{guzzo14} and \citet{garilli14}. Here we briefly summarize the last part of this process, i.e. how redshifts are measured and their quality evaluated.

As the final step of the VIPERS automated data-reduction pipeline developed at INAF--IASF Milano \citep{garilli14}, redshifts were estimated using the {\tt EZ} code \citep{garilli10}.  These measurements are then reviewed and confirmed or modified by two team members independently, using {\tt EZ} in interactive mode through a user-friendly dedicated interface. The results of the two reviews are eventually matched, and differences reconciled, to produce the final redshift estimate and its associated quality flag. 

All redshift measurements presented in the PDR-2 catalogue are as observed and have not been corrected to a heliocentric or Local-Group reference frame. Information to perform these corrections is nevertheless contained in the FITS header of the spectra.

\subsection{Redshift quality flags}
\label{sec:redshift_flags}

The quality flag system adopted by the VIPERS survey has been inspired by and is in fact very close to those of other precursor surveys \citep[e.g.][]{lefevre05,lilly09}. The meaning of the various flags has been described in detail in \citet{garilli14} and \citet{guzzo14}; here we repeat the meaning of the flags for those objects released with PDR-2:

\begin{itemize} 

\item
    Flags 4.X and 3.X: highly secure redshift, with confidence $>99\%$ 
\item
    Flag 2.X: still fairly secure, $> 95\%$ confidence level  
\item
   Flag 1.X: tentative redshift measurement, with $\sim 50\%$ chance to be wrong
\item
    Flag 9.X: redshift based a single emission feature, usually [OII]3727 \AA. With the PDR-1 data we showed that the confidence level of this class is $\sim 90\%$
    
\end{itemize}

After the human validation procedure has produced the integer part of the redshift quality flag, a decimal fraction is added to it, with possible values 0.2, 0.4, 0.5, to indicate respectively no, marginal or good agreement of the spectroscopic measurement with the object photometric redshift (see \citealt{guzzo14} for the specific criteria defining this agreement). If no photometric redshift exists for that object, the decimal part is set to 0.1.

A "1" in front of the above flags indicates a broad lines AGN spectrum, while a "2" indicates a second object serendipitously observed within the slit of a VIPERS target.

   \begin{figure}
   \centering
   \includegraphics[width=\hsize]{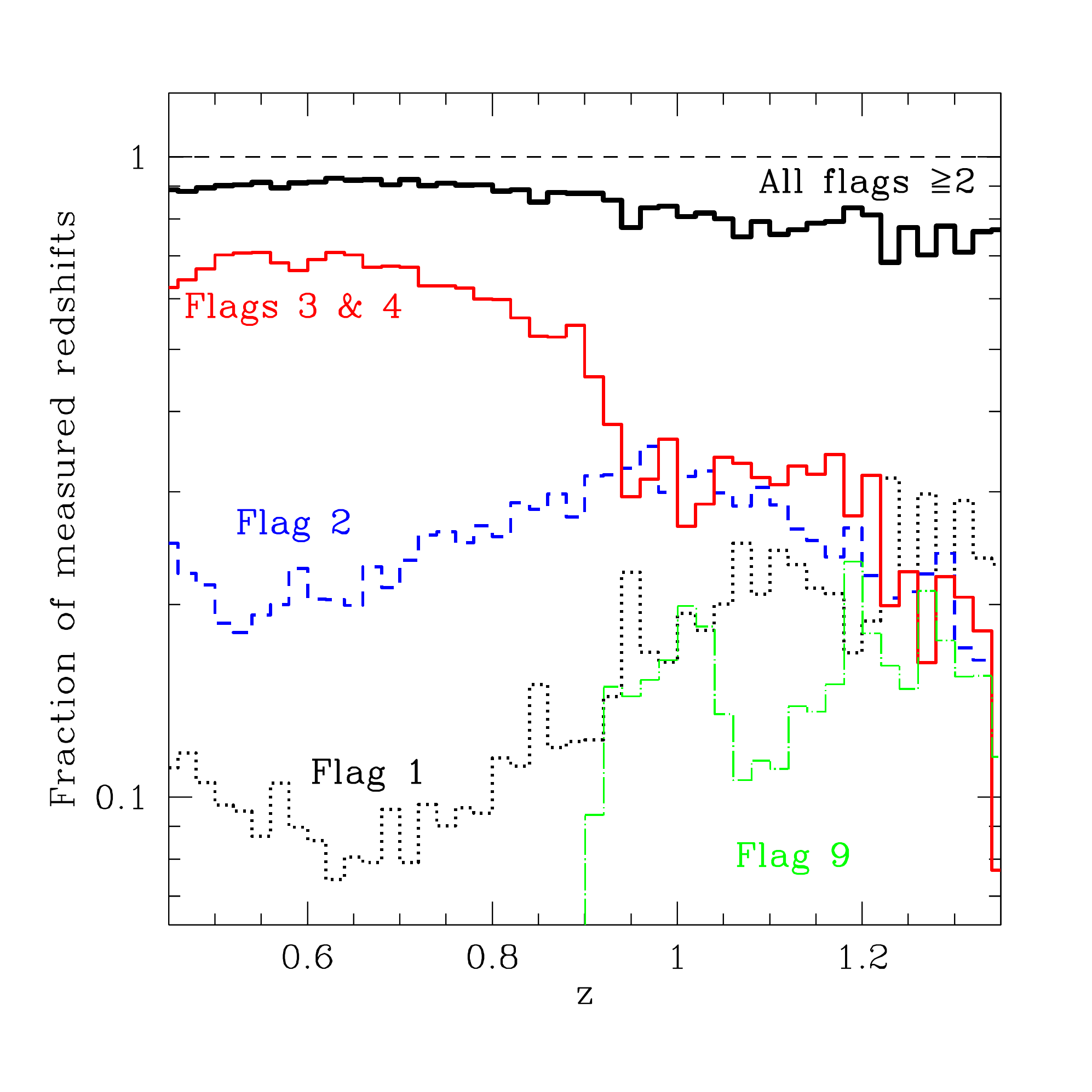}
     \caption{Quality of the VIPERS redshift measurements for different redshifts. Specifically, the plot shows how the fraction of measurements for different quality flags changes as a function of redshift.  Note how the "reliable sample", i.e. that with quality flag $\ge 2$, to be used for statistical analyses, shows a stable measured fraction out to the limit of the survey. 
              }
         \label{fig:flag_fractions}
   \end{figure}

In all VIPERS papers, objects with a redshift flag between 2.X and 9.X are referred to as {\sl reliable (or secure) redshifts} and are the only ones normally used in the science analyses.  In \citet{garilli14} we discussed in detail the reliability of flag 9.X objects.
In Fig.~\ref{fig:flag_fractions} the fraction of redshift measurements with a given quality flag is shown as a function of redshift, limited to the main redshift range covered by the VIPERS sample. Notice how the highest quality subset of redshift measurements is always the largest one in the survey, up to $z\simeq 1.2$ and also how, very importantly, the fraction of measured redshifts with flag 2.X or larger (i.e. the reliable sample to be used for science investigations) is essentially constant out to at least $z=1.2$.

%
\begin{table*}
\caption{Statistics of double redshift measurements, both internal to the VIPERS sample and against external data. } 
\label{tab:redshift_compare}      
\centering          
\begin{tabular}{l c r c c c c c}     
\hline\hline    
\rule{0pt}{2.5ex} Sample & $N_{\rm gal}$ & $N_{\rm agree}$ & Matching & $\sigma_2(\Delta z/(1+z))$ & $\sigma_z = \sigma_2 /\sqrt{2}$ & $\sigma_v = c \sigma_z$  & Mean $\Delta v$ \rule[-1.2ex]{0pt}{0pt}\\
&&& (\%) &&& (km\,s$^{-1}$) & (km\,s$^{-1}$)\\

\hline
\multicolumn{6}{c}{\rule{0pt}{2.5ex} Internal Comparison} \rule[-1.2ex]{0pt}{0pt}\\
\hline
\rule{0pt}{2.5ex} All measurements & 3,114 & 2,626 & 84.3\% & 0.00079 & 0.00056 & 166 & --\\
Both flags $\geq 2$ & 2,466 & 2,275 & 92.3\% & 0.00077 & 0.00054 & 163 & -- \\
Both flags 3,4 & 1,249 & 1,238 & 99.1\% & 0.00075 & 0.00053 & 159 & --\\

Both flags $\geq 2$; faint & -- & 1,099 & -- & 0.00077 & 0.00055 & 164 & --\\
Both flags $\geq 2$; bright & -- & 336 & -- & 0.00071 & 0.00050 & 150 & --\\
Both flags $\geq 2$; weak EL & -- & 1,153 & -- & 0.00079 & 0.00056 & 167 & --\\
Both flags $\geq 2$; strong EL & -- & 508 & -- & 0.00066 & 0.00046 & 139 & --\\

\hline
\multicolumn{6}{c}{\rule{0pt}{2.5ex} External Comparison with VVDS}\\
\hline
\rule{0pt}{2.5ex} All measurements & 737 & 629 & 85.3\% & 0.00093 & 0.00066 & 198 & 26 \\
Both flags 3,4 & 358 & 350 & 97.8\% & 0.00083 & 0.00059 & 177 & 40 \\

\hline
\multicolumn{6}{c}{\rule{0pt}{2.5ex} External Comparison with BOSS}\\
\hline
\rule{0pt}{2.5ex} All measurements & 747 & 736 & 98.5\% & 0.00064 & 0.00045* & 136* & $-108$\\
VIPERS flags 3,4 & 690 & 684 & 99.1\% & 0.00061 & 0.00043* & 130* & $-113$ \\

\hline\hline    
                
\end{tabular}

\vspace{1ex}
\raggedright{\rule{0pt}{2.5ex}* This is probably an underestimate of the uncertainty on the VIPERS side, because the assumption of uncertainty equipartition does not fully apply in the BOSS comparison}
\end{table*}
%

\subsection{Updated estimate of redshift reliability}  
\label{sec:redshift_reliability}

As was done for PDR-1, we estimate redshift errors by comparing independent redshift measurements that are available for a subset of galaxies. Some VIPERS targets were observed more than once within the survey, or are in common with other surveys. This also gives us a way to quantify the confidence level of our quality flags.  

At the end of the survey, the total number of targets with repeated observations is 3,556, compared to 1,941 that were available at the time of PDR-1. For 3114 of these, two redshift measurements are available, including any value for the quality flag (see Table~\ref{tab:redshift_compare}). 

Considering the distribution of the differences between the two redshift measurements $\Delta z$, we define matching pairs as those that satisfy the condition $|\Delta z| < 0.005$. This threshold has been set on the basis of the first visible gap in the $\Delta z$ distribution. This identifies 2,626 matching pairs (i.e. a matching fraction of 84\%).  This sample still includes some redshifts with quality flag 1, i.e. redshifts with confidence level $\sim 50-60 \%$ \citep[see][]{guzzo14}, which are in general not reliable for statistical analyses and have been excluded from all VIPERS investigations so far. Restricting consideration to pairs with both flag 2.X or above (i.e the reliable measurements), the matching fraction rises to 92.3\% (2,275 out of 2,466). Even further, if we consider only flags either 3.X or 4.X, i.e. the highest quality spectra, the matching reaches 99.1\% (1,238 out of 1,249). Using eq.~(7) in \citet{garilli14}, we can employ these figures to estimate the average confidence level of single measurements in the reliable redshift sample
\begin{equation}
C_{\rm flag \ge 2} = \sqrt{0.923} = 96.1\% \,\,\, ,
\end{equation}
and the one of the high-quality redshifts
\begin{equation}
C_{\rm flag3,4} = \sqrt{0.991} = 99.54\% \,\,\, ,
\end{equation}
which agrees very well with the value obtained in \citet{guzzo14} and \citet{garilli14} using the PDR-1 data. 

Comparable results are obtained when comparing the VIPERS measurements to external data. The VIPERS sky areas have a non negligible overlap with the VVDS Wide F22 sample \citet[844 galaxies]{garilli08} and the BOSS sample \citep[751 galaxies]{dawson13}. 
The results of these comparisons are shown in Table~\ref{tab:redshift_compare}. Note the significantly higher matching fraction in the case of BOSS. This is easily understood when considering that BOSS CMASS galaxies have a magnitude limit brighter than $i \simeq 20$ and so all matches must correspond to bright VIPERS galaxies. Not surprisingly, then, the matching fraction in this case is comparable to that of the highest quality VIPERS class. 

   \begin{figure}
   \centering
   \includegraphics[width=\hsize]{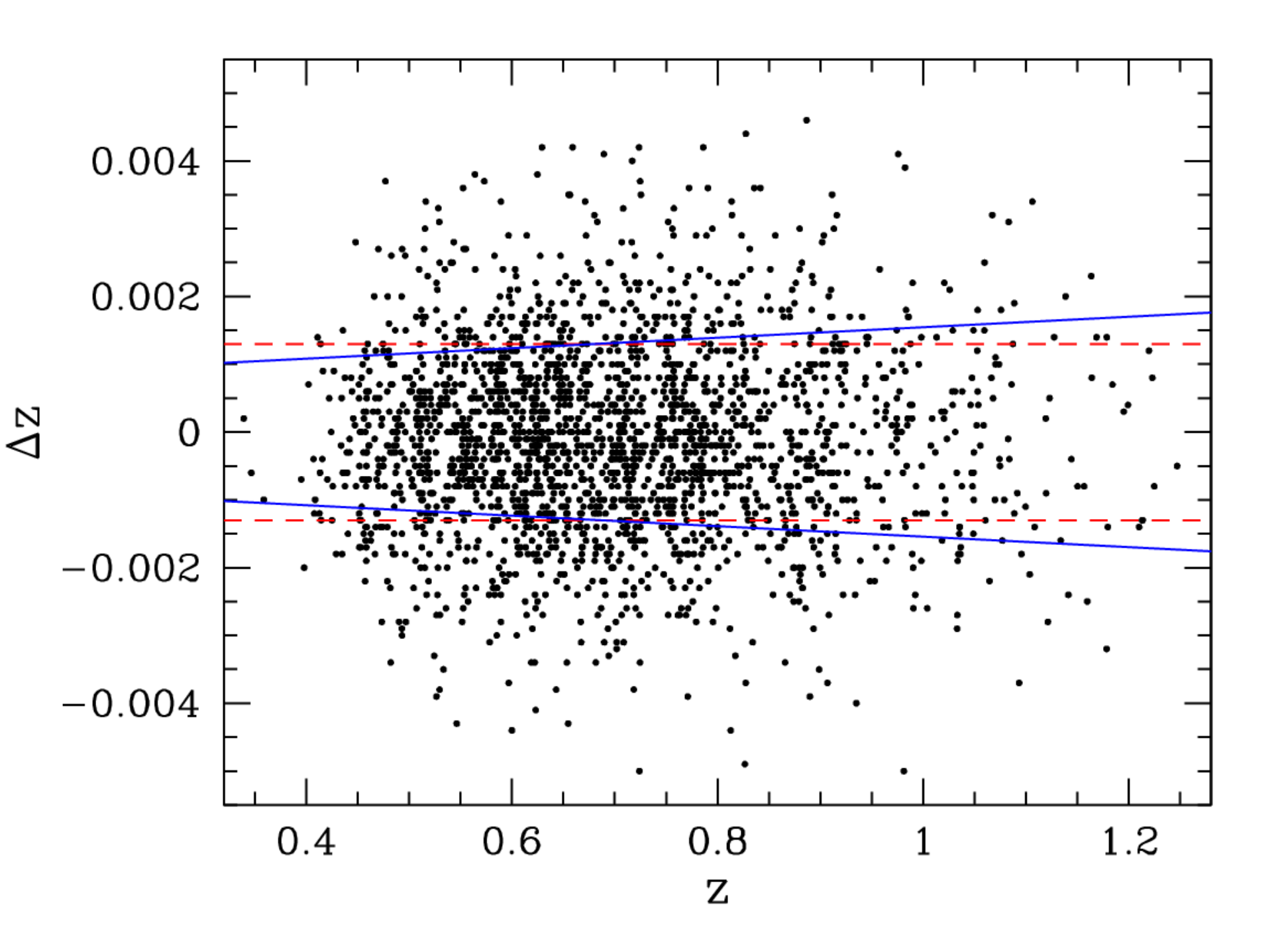}
     \caption{Redshift measurement differences
     between two independent observations of the same object, obtained
       from a set of 2,466 VIPERS galaxies with quality flag $\ge 2$, plotted as a function of the object redshift. Non-matching measurements (defined as being discrepant by more than $|\Delta z| = 0.005$) have been excluded from the plot. The dashed red and continuous blue lines show the 1-sigma scatter estimates obtained assuming the scatter to be independent from z, or scaling as $(1+z)$, respectively.
       }
         \label{fig:double-obs}
   \end{figure}

\subsection{Updated estimate of redshift errors}
\label{sec:redshift_errors}

An accurate knowledge of the typical redshift measurement uncertainty is clearly important for the scientific analysis of the VIPERS sample, in particular when modelling the observed shape of the power spectrum or the effect of Redshift Space Distortions \citep[see the parallel papers by][]{rota16, delatorre16, pezzotta16, wilson16}. It is also important to probe the possible redshift dependence of this uncertainty. We have used the repeated redshift measurements discussed in the previous section to update the estimate originally presented in \citet{guzzo14} and \citet{garilli14}.
To obtain a robust estimate of the measurement error we used the matching pairs of measurements (as defined in the previous section), computed the median absolute deviation (${\rm MAD}$) of the $\Delta z$ values, and scaled it to the standard deviation equivalent, which for a Gaussian distribution is given by  $\sigma = 1.4826 \times {\rm MAD}$.  The resulting scatter, when we consider only reliable redshift measurements (quality flag 2.X and above, 2,275 pairs), is $\sigma_{\Delta z} = 0.0013$. A very similar result is obtained by fitting a Gaussian to the distribution of $\Delta z$ values. 

With the current large set of duplicate measurements, we can also re-consider the overall approach used to characterise the redshift measurement uncertainty. For PDR-1 we adopted the common assumption that redshift uncertainties scale with the redshift itself as $(1+z)$. This assumption would apply in the simple case of a spectrograph that yields spectra with a resolution and sensitivity that are constant and independent of wavelength, but neither of these criteria are satisfied by VIMOS spectra.
The spectrum signal-to-noise ratio is influenced by the observing conditions, but is nevertheless mostly driven by the galaxy apparent magnitude. This in turn depends (albeit with a significant scatter) on the galaxy redshift, and therefore must induce some increase of the redshift measurement error with redshift. Also, when redshifts are measured through cross-correlation with templates, as in the case of VIPERS, some scaling of the uncertainty with redshift could be expected because of the $(1+z)$ shrinking of the available rest-frame wavelength range (although when the redshift measurement is dominated by a few key features in the spectrum, which remain observable over most of the survey redshift range -- e.g.~the ${\rm [OII]\lambda}3727$ line and \smash{4000\AA}~break region -- this effect should be negligible).  In short, a redshift dependence of the measurement error can be expected, but not necessarily with a linear dependence on $(1+z)$. 

   \begin{figure}
   \centering
   \includegraphics[width=\hsize]{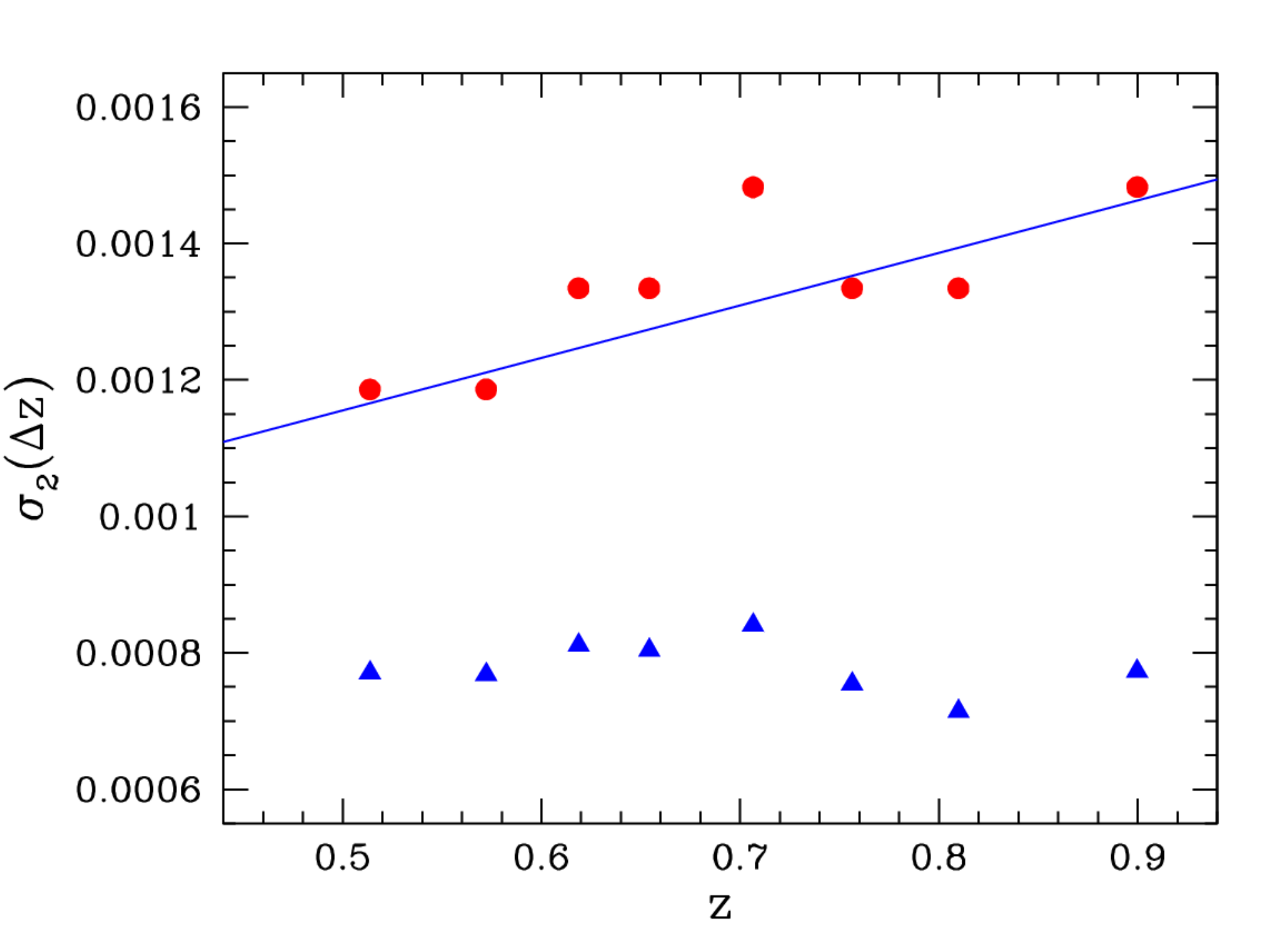}
     \caption{The scatter in the $\Delta z$ values (red circles)
     and in the $\Delta z / (1+z)$ values (blue triangles) for objects in a narrow
     redshift bin, as a function of the median redshift for the bin. 
     The bins are built to contain a fixed number of objects (500), and they are 
     not statistically independent since they partly overlap each other. 
     The blue line shows the mean scatter derived from the blue points, 
     with the $(1+z)$ scaling applied to it. 
     }
     \label{fig:scatter_vs_z}
   \end{figure}

The observed distribution of $\Delta z$ values as a function of galaxy redshift is shown in Fig.~\ref{fig:double-obs}, limited to the matching pairs of measurements. Using these data points, we have determined the scatter in both the $\Delta z$ and the $\Delta z / (1+z)$ values within separate redshift bins, still using the MAD estimator as for the whole sample. The results of this estimate are shown in Fig.~\ref{fig:scatter_vs_z}, where it is clearly seen that the observed scatter of the $\Delta z / (1+z)$ differences is substantially constant, whereas the scatter of the $\Delta z$ differences increases with the median redshift of the bin, with a scaling that is roughly proportional to $(1+z)$. We therefore conclude that this simple scaling does after all provide an adequate description of the effective uncertainty of the VIPERS redshift measurements. We do not extend this exercise to redshifts above 1.1, because of the limited number of repeated measurements available at those redshifts.
The final estimate we obtain for the single redshift measurement uncertainty is therefore $\sigma_z = 0.00054 \times (1+z)$, which we can compare with the figure of $\sigma_z = 0.00047 \times (1+z)$ given in the PDR-1 paper (on the basis of 1,235 measurement pairs). 
Notice that this is the estimate for the single measurement uncertainty, which we derive from the scatter measurement listed in Table~\ref{tab:redshift_compare} by assuming an equal contribution from the two measurements to the overall uncertainty, and therefore the single measurement uncertainty is obtained as the scatter value divided by a factor of $\smash{\sqrt{2}}$.

Using the sample of 2,275 reliable pairs of measurements (flags 2, 3, 4, 9), it is interesting to further explore how this scatter varies with the galaxy apparent magnitude or when strong emission lines are present in the spectrum. We isolate bright and faint subsets (galaxies with $i_{\rm AB} \leq 20.75$ and $i_{\rm AB} \geq 21.75$, respectively), and subsets with strong or weak emission-line spectra (galaxies with flux([OII]) $\geq 1.5\times 10^{-16}$ and between $2.0\times 10^{-17}$ and $1.5\times 10^{-16}$ $erg\ cm^{-2} s^{-1} A^{-1}$, respectively). As expected, significant differences are observed between these subsets. Table~\ref{tab:redshift_compare} shows that galaxies which are brighter or have strong emission lines in their spectra provide on average significantly more accurate redshift measurements, respectively by about 10 and 20\%.  

As in the case of the confidence level estimates, the comparison with independent surveys further confirms the accuracy of the VIPERS redshift measurements. The scatter between VIPERS and the VVDS is only marginally larger than that observed in the internal comparison. The scatter against BOSS is instead significantly smaller, because the higher spectral resolution of the BOSS spectra and the brighter magnitude limit of the BOSS sample both contribute to a higher redshift measurement accuracy for the BOSS galaxies, and therefore in this case we do not have equal contributions to the observed scatter, with the result that this is somewhat reduced in its observed value.

\section{The PDR-2 dataset}
\label{sec:dataset}

%
\begin{table}
\caption{The VIPERS PDR-2 spectroscopic sample} 
\label{tab:pdr2_numbers}      
\centering          
\begin{tabular}{l r}     
\hline\hline  

\rule{0pt}{2.5ex} Sample & Number\rule[-1.2ex]{0pt}{0pt}\\

\hline
\rule{0pt}{2.5ex} \noindent Spectroscopically observed & 97,414 \\
--- Main survey targets & 94,335 \\
--- Serendipitous targets & 1,478 \\
--- AGN candidates (not part & 1,601 \\
of main survey)\\

\hline
\hline
\rule{0pt}{2.5ex} Measured redshifts & Number\rule[-1.2ex]{0pt}{0pt}\\
\hline

\rule{0pt}{2.5ex} All measured & 91,507\\
Main survey, all targets & 89,022 \\
--- galaxies & 86,775 \\
--- stars & 2,247 \\
Flag $\geq 2$ main survey, all targets & 78,586\\
Flag $\geq 2$ main survey, galaxies & 76,552\rule[-1.2ex]{0pt}{0pt}\\

%
\hline\hline    
                
\end{tabular}
\end{table}
%

   \begin{figure}
   \centering
   \includegraphics[width=\hsize]{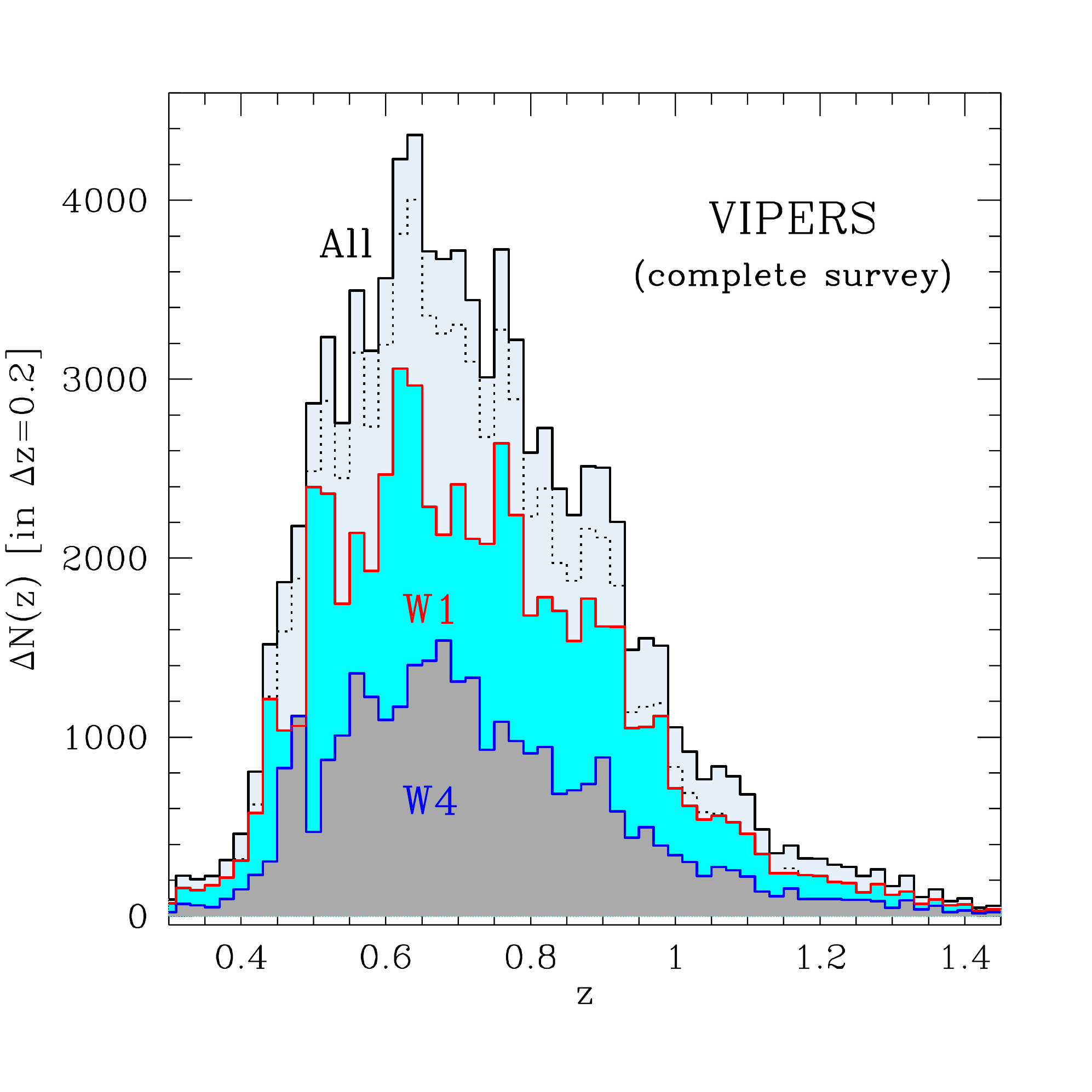}
      \caption{The galaxy redshift distribution in the final VIPERS
        PDR-2 catalogue (black solid line), and separately within the
        W1 and W4 fields (red and blue solid lines, respectively). These include all measured redshifts, with flag 1 or larger. The dotted line shows the result of plotting only the flag $\ge 2$ galaxies, i.e. those that can be used reliably in statistical analyses. 
              }
         \label{fig:nz}
   \end{figure}

\subsection{The PDR-2 sample}

In total we have obtained spectra for 97,414 objects: 94,335 main survey targets (i.e. the objects selected on the basis of the colour-colour selection criterion discussed in Sect.~\ref{sec:design}), 1,478 serendipitous targets in the slits, and 1,601 AGN candidates, originally selected on the basis of their colour. The total number of measured redshifts is 91,507 (quality flags 1 and above); of these, 86,775  make up the main galaxy survey, with the remaining ones belonging to stars and colour selected AGN. The galaxy sample with reliable redshift measurements (quality flag from 2.X to 9.X), contains 76,552 objects, with a median redshift of 0.69, and with 90\% of the objects located within the redshift range (0.43, 1.04). 

\begin{figure*}
   \includegraphics[width=9cm]{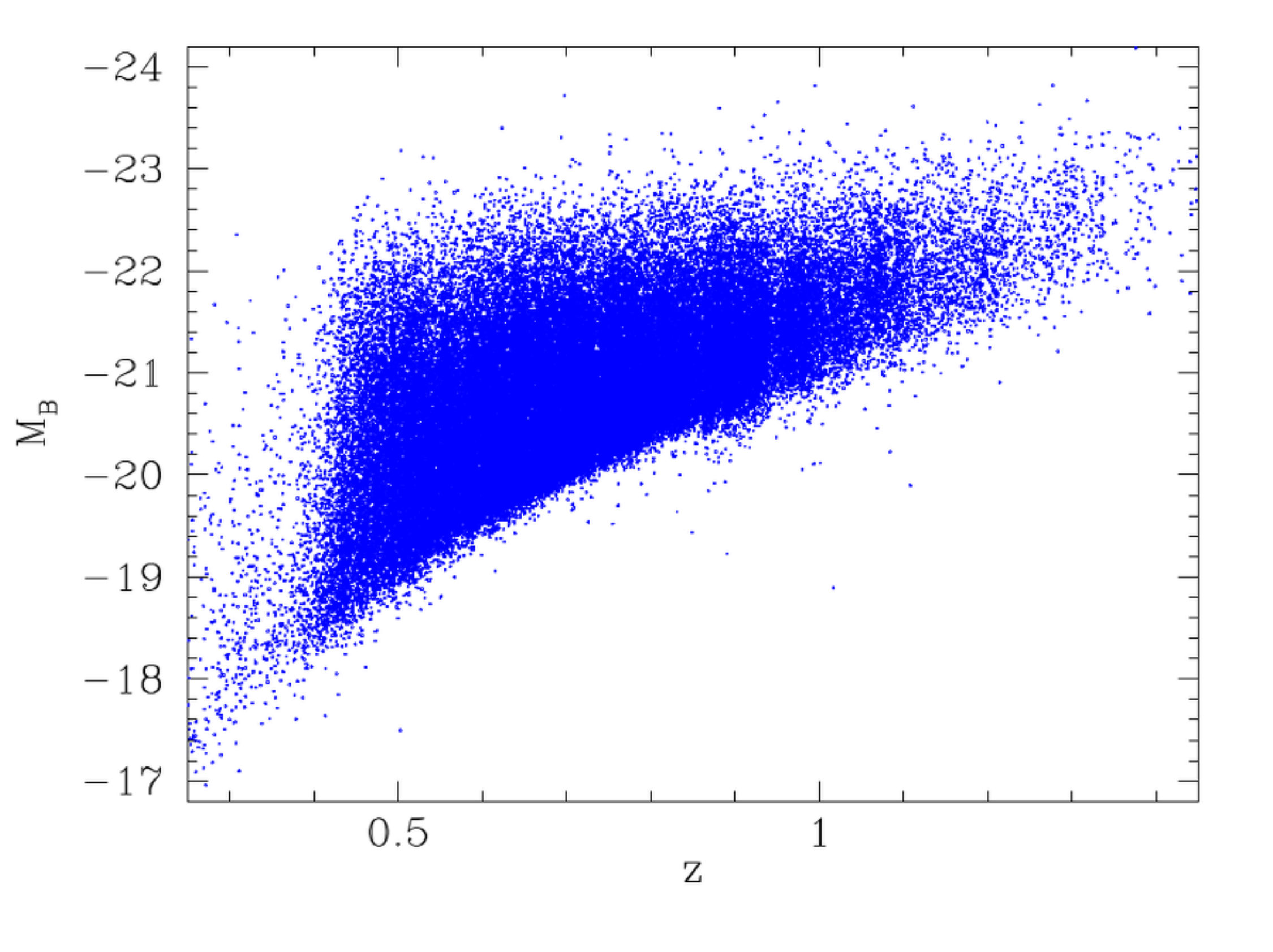}
   \includegraphics[width=9cm]{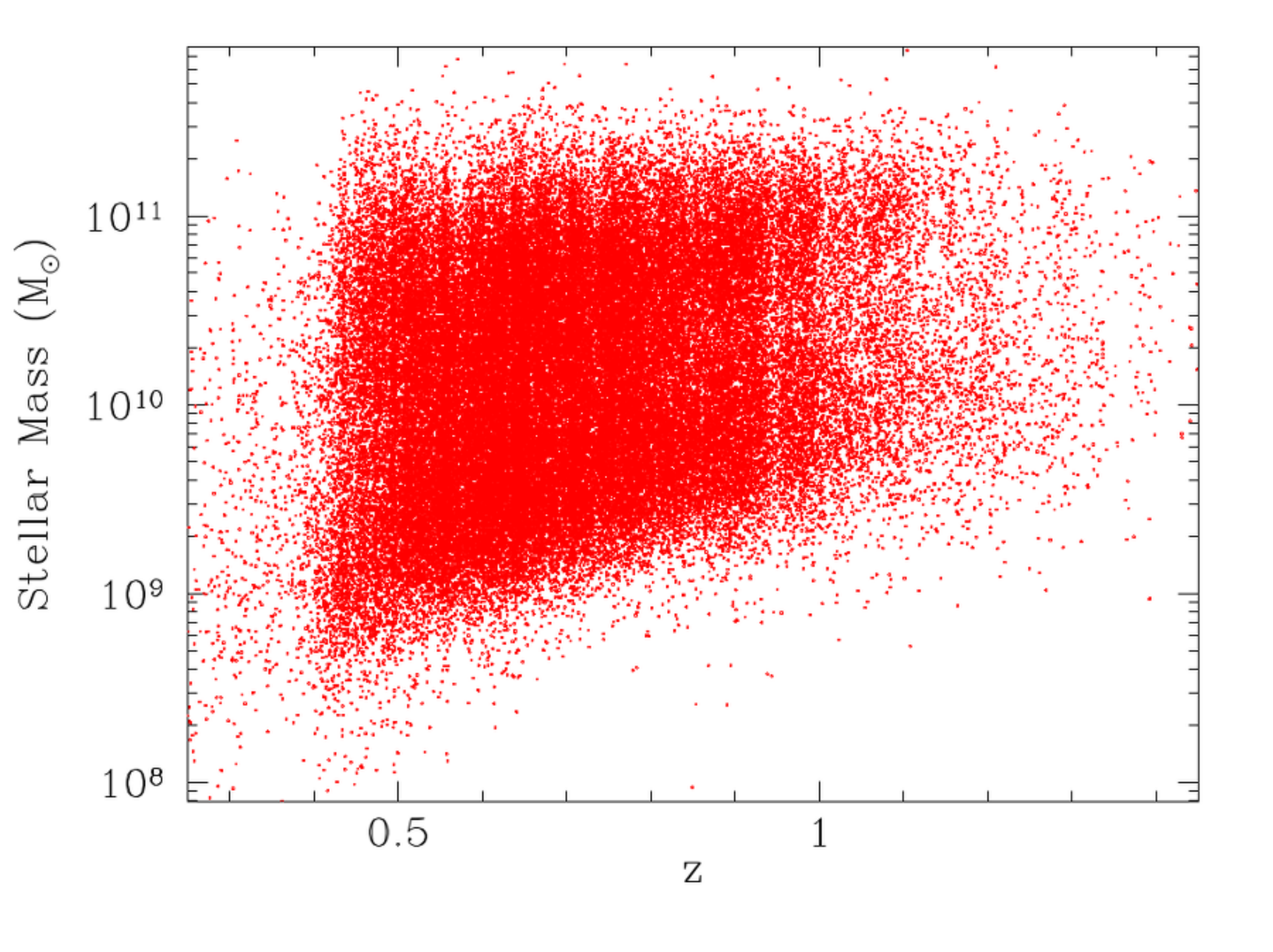}
      \caption{The distribution of $B$-band luminosities (left) and stellar masses (right) computed for all objects with reliable redshift (flag $>2.0$) in the PDR-2 catalogue. These are estimated through SED fitting of all available photometry (from UV to $K$) as described in \citet{moutard16b}. Note that an explicit value of $H_0=70\,{\rm km}\,{\rm s^{-1}}\,{\rm Mpc^{-1}} $ for the Hubble constant is used here. 
              }
         \label{fig:MB_z}
   \end{figure*}

Fig.~\ref{fig:nz} shows the redshift distribution of the final data set, providing also the two separate distributions for the two survey areas W1 and W4. This gives a visual impression of the level  of sample variance still present in the redshift distribution averaged over areas of this size. The two panels of Fig.~\ref{fig:MB_z} show instead the distribution of $B$-band luminosity and of stellar mass for the reliable redshift sample, to give a quantitative idea of the distribution of these two important galaxy properties within the VIPERS sample. Luminosities and stellar masses have been estimated through SED fitting of the available photometry (from UV to $K$) as described in \citet{moutard16b}. It is clear from these figures that the colour-colour selection used to select the VIPERS parent sample (see Sect.~\ref{sec:design}) has been highly effective in selecting a well defined galaxy sample at $z > 0.5$, which includes a sizeable set of very massive and bright galaxies.  

\begin{figure*}
   \centering
   \includegraphics[width=18cm]{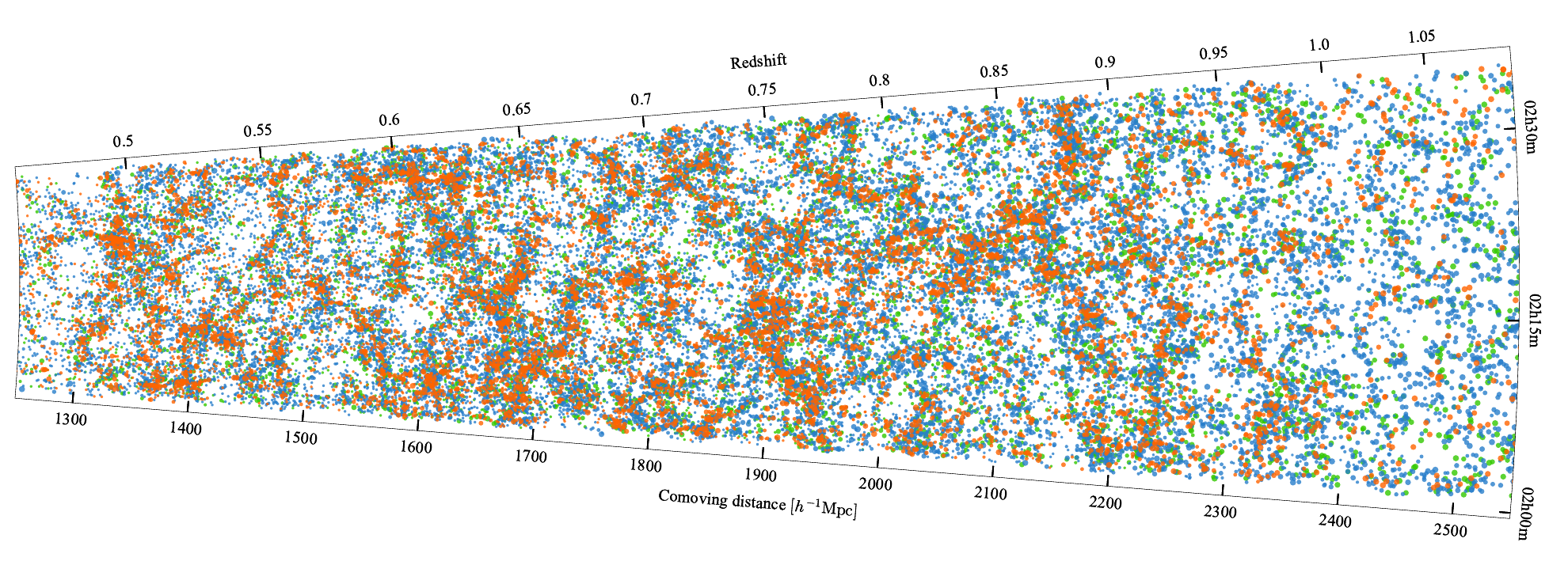}
   \includegraphics[width=18cm]{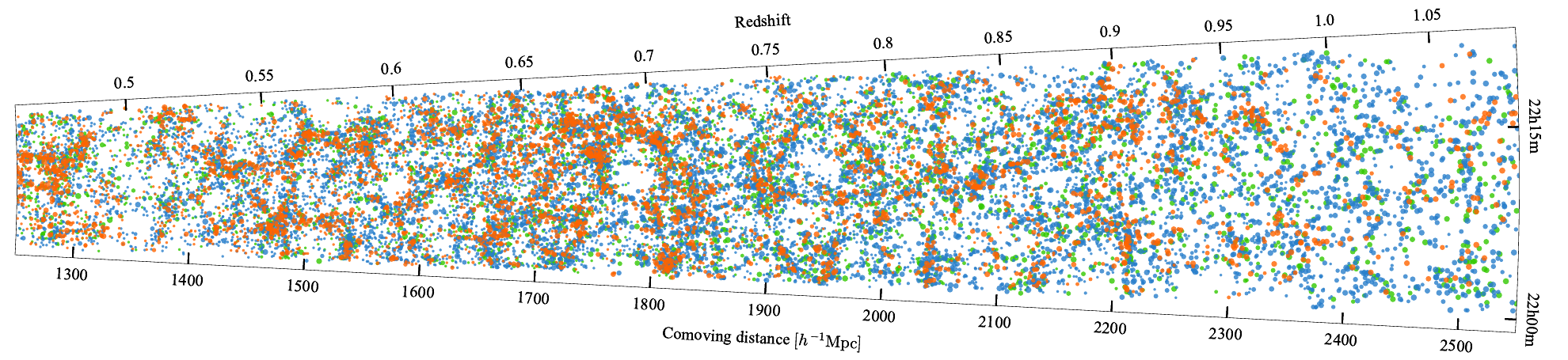}
     \caption{The distribution of galaxies at $0.45<z<1.1$ as shown 
     by the final VIPERS PDR-2 catalogue in the W1 and W4 fields (top and bottom cone diagram, respectively).  Galaxy positions are projected along declination (1.5 degrees for W4, about 1.8 degrees for W1). Each galaxy is represented by a filled circle of size proportional to its $B$-band luminosity and coloured according to its rest-frame $U-V$ colour.
              }
         \label{fig:coneW1W4}
   \end{figure*}

Some detail about the large-scale distribution of the VIPERS galaxies is instead provided by the cone diagrams of Fig.~\ref{fig:coneW1W4}, which show quite clearly the abundance of structures sampled by VIPERS, and the amount of segregation of the overall galaxy population as a function of the local galaxy density. 

\subsection{Access to publicly released data}

The PDR-2 data release includes: 
\begin{itemize}
\item 
the total reference photometric catalogue, including both the VIPERS parent photometric sample, the objects excluded from this sample on the basis of the colour-colour criterion described in Sect.~\ref{sec:design}, and the objects classified as stars; the parameter classFlag is used in the catalogue to identify these categories (objects in the VIPERS parent photometric sample have classFlag=1).
The catalogue contains both the CFHTLS release T0005 photometry used for the VIPERS sample selection (see Sect.~\ref{sec:design}), and the VIPERS Multi-Lambda photometry (CFHTLS release T0007 supplemented with UV and $K$-band data), described in Sect.~\ref{sec:cfhtls_t07}. 
\item
the photometric and spectroscopic masks described in Sect.~\ref{sec:photo_mask} and \ref{sec:spectro_mask}, in the form of DS9 region files.
\item 
the redshift catalogue with the parameters described in Sect.~\ref{sec:redshift_measure}, and which includes the estimates of the Target Sampling Rate and Spectroscopic Success Rate described in Sect.~\ref{sec:tsr} and \ref{sec:ssr}. This catalogue includes all the available redshift measurements (quality flag $\geq 1$), and not only those in the reliable redshift sample. Also in this catalogue the parameter classFlag can be used to identify the objects that belong to the main survey sample (objects in this sample have classFlag=1).
\item
the wavelength- and flux-calibrated one-dimensional spectra for the objects included in the redshift catalog; these are provided as FITS tables containing the following columns:
1) wavelength (in Angstrom);
2) cleaned-spectrum flux (in $erg\ cm^{-2} s^{-1} A^{-1}$); the spectral cleaning procedure, based on a PCA reconstruction of the observed spectra to remove the strongest residuals from the sky subtraction, is described in detail in \cite{marchetti16}; fluxes are also corrected for atmospheric absorption effects;
3) flux uncertainty estimate (same units as the flux);
4) the sky intensity estimate for the slit (in counts);
5) original spectrum flux (same units as the flux), with only the correction for atmospheric absorption applied;
6) the cleaned-spectrum mask (digital mask to identify the pixels edited by the cleaning procedure).
\item
optionally, also the two-dimensional spectra are available; these are wavelength calibrated, but without the flux calibration nor the correction for atmospheric absorption applied.
\end{itemize}

The complete data release of VIPERS is available at {\tt http://vipers.inaf.it}, while the spectroscopic part is also available via the ESO archive at {\tt http://eso.org/rm/publicAccess\#/dataReleases}.

\begin{acknowledgements}

      We acknowledge the crucial contribution of the ESO staff for the management of service observations. In particular, we are deeply 
      grateful to M. Hilker for his constant help and support of this program. Italian participation to VIPERS has been funded by INAF 
      through PRIN 2008, 2010, and 2014 programs. LG, JB and BRG acknowledge support from the European Research Council through grant n.~291521. 
      OLF acknowledges support from the European Research Council through grant n.~268107. JAP acknowledges support of the European Research Council 
      through grant n.~67093. WJP and RT acknowledge financial support from the European Research Council through grant n.~202686. AP, KM, and JK have 
      been supported by the National Science Centre (grants UMO-2012/07/B/ST9/04425 and UMO-2013/09/D/ST9/04030). WJP is also grateful for support from 
      the UK Science and Technology Facilities Council through the grant ST/I001204/1. EB, FM and LM acknowledge the support from grants ASI-INAF 
      I/023/12/0 and PRIN MIUR 2010-2011. LM also acknowledges financial support from PRIN INAF 2012. YM acknowledges support from CNRS/INSU 
      (Institut National des Sciences de l’Univers). TM and SA  acknowledge financial support from the ANR Spin(e) through the french grant  ANR-13-BS05-0005. CM is grateful for support from specific project funding of the {\it Institut Universitaire 
      de France}. SDLT and CM acknowledge the support of the OCEVU Labex (ANR-11-LABX-0060) and the A*MIDEX project (ANR-11-IDEX-0001-02) funded 
      by the "Investissements d'Avenir" French government program managed by the ANR. and the Programme National Galaxies et Cosmologie (PNCG). 
      Research conducted within the scope of the HECOLS International Associated Laboratory, supported in part by the Polish NCN grant DEC-2013/08/M/ST9/00664.

\end{acknowledgements}

\bibliographystyle{aa}
\bibliography{biblio_gg,biblio_VIPERS_v2}

\begin{thebibliography}{32}
\expandafter\ifx\csname natexlab\endcsname\relax\def\natexlab#1{#1}\fi

\bibitem[{{Abazajian} {et~al.}(2009){Abazajian}, {Adelman-McCarthy},
  {Ag{\"u}eros}, {Allam}, {Allende Prieto}, {An}, {Anderson}, {Anderson},
  {Annis}, {Bahcall}, \& et~al.}]{sdss_dr7}
{Abazajian}, K.~N., {Adelman-McCarthy}, J.~K., {Ag{\"u}eros}, M.~A., {et~al.}
  2009, \apjs, 182, 543

\bibitem[{{Ahn} {et~al.}(2012){Ahn}, {Alexandroff}, {Allende Prieto},
  {Anderson}, {Anderton}, {Andrews}, {Aubourg}, {Bailey}, {Balbinot}, {Barnes},
  \& et~al.}]{sdss_dr9}
{Ahn}, C.~P., {Alexandroff}, R., {Allende Prieto}, C., {et~al.} 2012, \apjs,
  203, 21

\bibitem[{{Alam} {et~al.}(2015){Alam}, {Albareti}, {Allende Prieto}, {Anders},
  {Anderson}, {Anderton}, {Andrews}, {Armengaud}, {Aubourg}, {Bailey}, \&
  et~al.}]{alam15}
{Alam}, S., {Albareti}, F.~D., {Allende Prieto}, C., {et~al.} 2015, \apjs, 219,
  12

\bibitem[{{Bottini} {et~al.}(2005){Bottini}, {Garilli}, {Maccagni}, {Tresse},
  {Le Brun}, {Le F{\`e}vre}, {Picat}, {Scaramella}, {Scodeggio}, {Vettolani},
  {Zanichelli}, {Adami}, {Arnaboldi}, {Arnouts}, {Bardelli}, {Bolzonella},
  {Cappi}, {Charlot}, {Ciliegi}, {Contini}, {Foucaud}, {Franzetti}, {Guzzo},
  {Ilbert}, {Iovino}, {McCracken}, {Marano}, {Marinoni}, {Mathez}, {Mazure},
  {Meneux}, {Merighi}, {Paltani}, {Pollo}, {Pozzetti}, {Radovich}, {Zamorani},
  \& {Zucca}}]{bottini05}
{Bottini}, D., {Garilli}, B., {Maccagni}, D., {et~al.} 2005, \pasp, 117, 996

\bibitem[{{Colless} {et~al.}(2003){Colless}, {Peterson}, {Jackson}, {Peacock},
  {Cole}, {Norberg}, {Baldry}, {Baugh}, {Bland-Hawthorn}, {Bridges}, {Cannon},
  {Collins}, {Couch}, {Cross}, {Dalton}, {De Propris}, {Driver}, {Efstathiou},
  {Ellis}, {Frenk}, {Glazebrook}, {Lahav}, {Lewis}, {Lumsden}, {Maddox},
  {Madgwick}, {Sutherland}, \& {Taylor}}]{colless03}
{Colless}, M., {Peterson}, B.~A., {Jackson}, C., {et~al.} 2003, ArXiv e-print,
  astro-ph/0306581

\bibitem[{{Dawson} {et~al.}(2013){Dawson}, {Schlegel}, {Ahn}, {Anderson},
  {Aubourg}, {Bailey}, {Barkhouser}, {Bautista}, {Beifiori}, {Berlind},
  {Bhardwaj}, {Bizyaev}, {Blake}, {Blanton}, {Blomqvist}, {Bolton}, {Borde},
  {Bovy}, {Brandt}, {Brewington}, {Brinkmann}, {Brown}, {Brownstein}, {Bundy},
  {Busca}, {Carithers}, {Carnero}, {Carr}, {Chen}, {Comparat}, {Connolly},
  {Cope}, {Croft}, {Cuesta}, {da Costa}, {Davenport}, {Delubac}, {de Putter},
  {Dhital}, {Ealet}, {Ebelke}, {Eisenstein}, {Escoffier}, {Fan}, {Filiz Ak},
  {Finley}, {Font-Ribera}, {G{\'e}nova-Santos}, {Gunn}, {Guo}, {Haggard},
  {Hall}, {Hamilton}, {Harris}, {Harris}, {Ho}, {Hogg}, {Holder}, {Honscheid},
  {Huehnerhoff}, {Jordan}, {Jordan}, {Kauffmann}, {Kazin}, {Kirkby}, {Klaene},
  {Kneib}, {Le Goff}, {Lee}, {Long}, {Loomis}, {Lundgren}, {Lupton}, {Maia},
  {Makler}, {Malanushenko}, {Malanushenko}, {Mandelbaum}, {Manera}, {Maraston},
  {Margala}, {Masters}, {McBride}, {McDonald}, {McGreer}, {McMahon}, {Mena},
  {Miralda-Escud{\'e}}, {Montero-Dorta}, {Montesano}, {Muna}, {Myers},
  {Naugle}, {Nichol}, {Noterdaeme}, {Nuza}, {Olmstead}, {Oravetz}, {Oravetz},
  {Owen}, {Padmanabhan}, {Palanque-Delabrouille}, {Pan}, {Parejko},
  {P{\^a}ris}, {Percival}, {P{\'e}rez-Fournon}, {P{\'e}rez-R{\`a}fols},
  {Petitjean}, {Pfaffenberger}, {Pforr}, {Pieri}, {Prada}, {Price-Whelan},
  {Raddick}, {Rebolo}, {Rich}, {Richards}, {Rockosi}, {Roe}, {Ross}, {Ross},
  {Rossi}, {Rubi{\~n}o-Martin}, {Samushia}, {S{\'a}nchez}, {Sayres}, {Schmidt},
  {Schneider}, {Sc{\'o}ccola}, {Seo}, {Shelden}, {Sheldon}, {Shen}, {Shu},
  {Slosar}, {Smee}, {Snedden}, {Stauffer}, {Steele}, {Strauss}, {Streblyanska},
  {Suzuki}, {Swanson}, {Tal}, {Tanaka}, {Thomas}, {Tinker}, {Tojeiro},
  {Tremonti}, {Vargas Maga{\~n}a}, {Verde}, {Viel}, {Wake}, {Watson}, {Weaver},
  {Weinberg}, {Weiner}, {West}, {White}, {Wood-Vasey}, {Yeche}, {Zehavi},
  {Zhao}, \& {Zheng}}]{dawson13}
{Dawson}, K.~S., {Schlegel}, D.~J., {Ahn}, C.~P., {et~al.} 2013, \aj, 145, 10

\bibitem[{{de la Torre} {et~al.}(2013){de la Torre}, {Guzzo}, {Peacock},
  {Branchini}, {Iovino}, {Granett}, {Abbas}, {Adami}, {Arnouts}, {Bel},
  {Bolzonella}, {Bottini}, {Cappi}, {Coupon}, {Cucciati}, {Davidzon}, {De
  Lucia}, {Fritz}, {Franzetti}, {Fumana}, {Garilli}, {Ilbert}, {Krywult}, {Le
  Brun}, {Le F{\`e}vre}, {Maccagni}, {Ma{\l}ek}, {Marulli}, {McCracken},
  {Moscardini}, {Paioro}, {Percival}, {Polletta}, {Pollo}, {Schlagenhaufer},
  {Scodeggio}, {Tasca}, {Tojeiro}, {Vergani}, {Zanichelli}, {Burden}, {Di
  Porto}, {Marchetti}, {Marinoni}, {Mellier}, {Monaco}, {Nichol}, {Phleps},
  {Wolk}, \& {Zamorani}}]{delatorre13}
{de la Torre}, S., {Guzzo}, L., {Peacock}, J.~A., {et~al.} 2013, \aap, 557, A54

\bibitem[{{de la Torre} {et~al.}(2016)}]{delatorre16}
{de la Torre}, S. {et~al.} 2016, \aap~\rm{submitted}, \rm{ArXiv e-print
  XXXX.YYYY}

\bibitem[{{Eisenstein} {et~al.}(2011){Eisenstein}, {Weinberg}, {Agol},
  {Aihara}, {Allende Prieto}, {Anderson}, {Arns}, {Aubourg}, {Bailey},
  {Balbinot}, \& et~al.}]{eisenstein11}
{Eisenstein}, D.~J., {Weinberg}, D.~H., {Agol}, E., {et~al.} 2011, \aj, 142, 72

\bibitem[{{Emerson} {et~al.}(2004){Emerson}, {Sutherland}, {McPherson},
  {Craig}, {Dalton}, \& {Ward}}]{Emerson2004}
{Emerson}, J.~P., {Sutherland}, W.~J., {McPherson}, A.~M., {et~al.} 2004, The
  Messenger, 117, 27

\bibitem[{{Garilli} {et~al.}(2010){Garilli}, {Fumana}, {Franzetti}, {Paioro},
  {Scodeggio}, {Le F{\`e}vre}, {Paltani}, \& {Scaramella}}]{garilli10}
{Garilli}, B., {Fumana}, M., {Franzetti}, P., {et~al.} 2010, \pasp, 122, 827

\bibitem[{{Garilli} {et~al.}(2014){Garilli}, {Guzzo}, {Scodeggio},
  {Bolzonella}, {Abbas}, {Adami}, {Arnouts}, {Bel}, {Bottini}, {Branchini},
  {Cappi}, {Coupon}, {Cucciati}, {Davidzon}, {De Lucia}, {de la Torre},
  {Franzetti}, {Fritz}, {Fumana}, {Granett}, {Ilbert}, {Iovino}, {Krywult}, {Le
  Brun}, {Le F{\`e}vre}, {Maccagni}, {Ma{\l}ek}, {Marulli}, {McCracken},
  {Paioro}, {Polletta}, {Pollo}, {Schlagenhaufer}, {Tasca}, {Tojeiro},
  {Vergani}, {Zamorani}, {Zanichelli}, {Burden}, {Di Porto}, {Marchetti},
  {Marinoni}, {Mellier}, {Moscardini}, {Nichol}, {Peacock}, {Percival},
  {Phleps}, \& {Wolk}}]{garilli14}
{Garilli}, B., {Guzzo}, L., {Scodeggio}, M., {et~al.} 2014, \aap, 562, A23

\bibitem[{{Garilli} {et~al.}(2008){Garilli}, {Le F{\`e}vre}, {Guzzo},
  {Maccagni}, {Le Brun}, {de la Torre}, {Meneux}, {Tresse}, {Franzetti},
  {Zamorani}, {Zanichelli}, {Gregorini}, {Vergani}, {Bottini}, {Scaramella},
  {Scodeggio}, {Vettolani}, {Adami}, {Arnouts}, {Bardelli}, {Bolzonella},
  {Cappi}, {Charlot}, {Ciliegi}, {Contini}, {Foucaud}, {Gavignaud}, {Ilbert},
  {Iovino}, {Lamareille}, {McCracken}, {Marano}, {Marinoni}, {Mazure},
  {Merighi}, {Paltani}, {Pell{\`o}}, {Pollo}, {Pozzetti}, {Radovich}, {Zucca},
  {Blaizot}, {Bongiorno}, {Cucciati}, {Mellier}, {Moreau}, \&
  {Paioro}}]{garilli08}
{Garilli}, B., {Le F{\`e}vre}, O., {Guzzo}, L., {et~al.} 2008, \aap, 486, 683

\bibitem[{{Garilli} {et~al.}(2012){Garilli}, {Paioro}, {Scodeggio},
  {Franzetti}, {Fumana}, \& {Guzzo}}]{garilli12}
{Garilli}, B., {Paioro}, L., {Scodeggio}, M., {et~al.} 2012, \pasp, 124, 1232

\bibitem[{{Guzzo} {et~al.}(2014){Guzzo}, {Scodeggio}, {Garilli}, {Granett},
  {Fritz}, {Abbas}, {Adami}, {Arnouts}, {Bel}, {Bolzonella}, {Bottini},
  {Branchini}, {Cappi}, {Coupon}, {Cucciati}, {Davidzon}, {De Lucia}, {de la
  Torre}, {Franzetti}, {Fumana}, {Hudelot}, {Ilbert}, {Iovino}, {Krywult}, {Le
  Brun}, {Le F{\`e}vre}, {Maccagni}, {Ma{\l}ek}, {Marulli}, {McCracken},
  {Paioro}, {Peacock}, {Polletta}, {Pollo}, {Schlagenhaufer}, {Tasca},
  {Tojeiro}, {Vergani}, {Zamorani}, {Zanichelli}, {Burden}, {Di Porto},
  {Marchetti}, {Marinoni}, {Mellier}, {Moscardini}, {Nichol}, {Percival},
  {Phleps}, \& {Wolk}}]{guzzo14}
{Guzzo}, L., {Scodeggio}, M., {Garilli}, B., {et~al.} 2014, \aap, 566, A108

\bibitem[{{Hammersley} {et~al.}(2010){Hammersley}, {Christensen}, {Dekker},
  {Izzo}, {Selman}, {Bristow}, {Bourget}, {Castillo}, {Downing}, {Haddad},
  {Hilker}, {Lizon}, {Lucuix}, {Mainieri}, {Mieske}, {Reinero}, {Rejkuba},
  {Rojas}, {Smette}, {Urrutia Del Rio}, {Valenzuela}, \&
  {Wolff}}]{hammersley10}
{Hammersley}, P., {Christensen}, L., {Dekker}, H., {et~al.} 2010, The
  Messenger, 142, 8

\bibitem[{{Hildebrandt} {et~al.}(2012){Hildebrandt}, {Erben}, {Kuijken}, {van
  Waerbeke}, {Heymans}, {Coupon}, {Benjamin}, {Bonnett}, {Fu}, {Hoekstra},
  {Kitching}, {Mellier}, {Miller}, {Velander}, {Hudson}, {Rowe}, {Schrabback},
  {Semboloni}, \& {Ben{\'{\i}}tez}}]{hildebrandt12}
{Hildebrandt}, H., {Erben}, T., {Kuijken}, K., {et~al.} 2012, \mnras, 421, 2355

\bibitem[{{Hudelot} {et~al.}(2012){Hudelot}, {Cuillandre}, {Withington},
  {Goranova}, {McCracken}, {Magnard}, {Mellier}, {Regnault}, {Betoule},
  {Aussel}, {Kavelaars}, {Fernique}, {Bonnarel}, {Ochsenbein}, \&
  {Ilbert}}]{hudelot12}
{Hudelot}, P., {Cuillandre}, J.-C., {Withington}, K., {et~al.} 2012, VizieR
  Online Data Catalog, 2317

\bibitem[{{Jarvis} {et~al.}(2013){Jarvis}, {Bonfield}, {Bruce}, {Geach},
  {McAlpine}, {McLure}, {Gonz{\'a}lez-Solares}, {Irwin}, {Lewis}, {Yoldas},
  {Andreon}, {Cross}, {Emerson}, {Dalton}, {Dunlop}, {Hodgkin}, {Le},
  {Karouzos}, {Meisenheimer}, {Oliver}, {Rawlings}, {Simpson}, {Smail},
  {Smith}, {Sullivan}, {Sutherland}, {White}, \& {Zwart}}]{Jarvis2013}
{Jarvis}, M.~J., {Bonfield}, D.~G., {Bruce}, V.~A., {et~al.} 2013, \mnras, 428,
  1281

\bibitem[{{Le F{\`e}vre} {et~al.}(2013){Le F{\`e}vre}, {Cassata}, {Cucciati},
  {Garilli}, {Ilbert}, {Le Brun}, {Maccagni}, {Moreau}, {Scodeggio}, {Tresse},
  {Zamorani}, {Adami}, {Arnouts}, {Bardelli}, {Bolzonella}, {Bondi},
  {Bongiorno}, {Bottini}, {Cappi}, {Charlot}, {Ciliegi}, {Contini}, {de la
  Torre}, {Foucaud}, {Franzetti}, {Gavignaud}, {Guzzo}, {Iovino}, {Lemaux},
  {L{\'o}pez-Sanjuan}, {McCracken}, {Marano}, {Marinoni}, {Mazure}, {Mellier},
  {Merighi}, {Merluzzi}, {Paltani}, {Pell{\`o}}, {Pollo}, {Pozzetti},
  {Scaramella}, {Tasca}, {Vergani}, {Vettolani}, {Zanichelli}, \&
  {Zucca}}]{lefevre13}
{Le F{\`e}vre}, O., {Cassata}, P., {Cucciati}, O., {et~al.} 2013, \aap, 559,
  A14

\bibitem[{{Le F{\`e}vre} {et~al.}(2003){Le F{\`e}vre}, {Saisse}, {Mancini},
  {Brau-Nogue}, {Caputi}, {Castinel}, {D'Odorico}, {Garilli}, {Kissler-Patig},
  {Lucuix}, {Mancini}, {Pauget}, {Sciarretta}, {Scodeggio}, {Tresse}, \&
  {Vettolani}}]{lefevre03}
{Le F{\`e}vre}, O., {Saisse}, M., {Mancini}, D., {et~al.} 2003, in \procspie,
  ed. M.~{Iye} \& A.~F.~M. {Moorwood}, Vol. 4841, 1670--1681

\bibitem[{{Le F{\`e}vre} {et~al.}(2005){Le F{\`e}vre}, {Vettolani}, {Garilli},
  {Tresse}, {Bottini}, {Le Brun}, {Maccagni}, {Picat}, {Scaramella},
  {Scodeggio}, {Zanichelli}, {Adami}, {Arnaboldi}, {Arnouts}, {Bardelli},
  {Bolzonella}, {Cappi}, {Charlot}, {Ciliegi}, {Contini}, {Foucaud},
  {Franzetti}, {Gavignaud}, {Guzzo}, {Ilbert}, {Iovino}, {McCracken}, {Marano},
  {Marinoni}, {Mathez}, {Mazure}, {Meneux}, {Merighi}, {Paltani}, {Pell{\`o}},
  {Pollo}, {Pozzetti}, {Radovich}, {Zamorani}, {Zucca}, {Bondi}, {Bongiorno},
  {Busarello}, {Lamareille}, {Mellier}, {Merluzzi}, {Ripepi}, \&
  {Rizzo}}]{lefevre05}
{Le F{\`e}vre}, O., {Vettolani}, G., {Garilli}, B., {et~al.} 2005, \aap, 439,
  845

\bibitem[{{Lilly} {et~al.}(2009){Lilly}, {Le Brun}, {Maier}, {Mainieri},
  {Mignoli}, {Scodeggio}, {Zamorani}, {Carollo}, {Contini}, {Kneib}, {Le
  F{\`e}vre}, {Renzini}, {Bardelli}, {Bolzonella}, {Bongiorno}, {Caputi},
  {Coppa}, {Cucciati}, {de la Torre}, {de Ravel}, {Franzetti}, {Garilli},
  {Iovino}, {Kampczyk}, {Kovac}, {Knobel}, {Lamareille}, {LeBorgne}, {Pello},
  {Peng}, {P{\'e}rez-Montero}, {Ricciardelli}, {Silverman}, {Tanaka}, {Tasca},
  {Tresse}, {Vergani}, {Zucca}, {Ilbert}, {Salvato}, {Oesch}, {Abbas},
  {Bottini}, {Capak}, {Cappi}, {Cassata}, {Cimatti}, {Elvis}, {Fumana},
  {Guzzo}, {Hasinger}, {Koekemoer}, {Leauthaud}, {Maccagni}, {Marinoni},
  {McCracken}, {Memeo}, {Meneux}, {Porciani}, {Pozzetti}, {Sanders},
  {Scaramella}, {Scarlata}, {Scoville}, {Shopbell}, \& {Taniguchi}}]{lilly09}
{Lilly}, S.~J., {Le Brun}, V., {Maier}, C., {et~al.} 2009, \apjs, 184, 218

\bibitem[{{Marchetti} {et~al.}(2016)}]{marchetti16}
{Marchetti}, A. {et~al.} 2016, \aap~\rm{submitted}, \rm{arXiv:XXXX.YYYY}

\bibitem[{{Martin} {et~al.}(2005){Martin}, {Fanson}, {Schiminovich},
  {Morrissey}, {Friedman}, {Barlow}, {Conrow}, {Grange}, {Jelinsky},
  {Milliard}, {Siegmund}, {Bianchi}, {Byun}, {Donas}, {Forster}, {Heckman},
  {Lee}, {Madore}, {Malina}, {Neff}, {Rich}, {Small}, {Surber}, {Szalay},
  {Welsh}, \& {Wyder}}]{Martin2005}
{Martin}, D.~C., {Fanson}, J., {Schiminovich}, D., {et~al.} 2005, \apjl, 619,
  L1

\bibitem[{{Moutard} {et~al.}(2016b){Moutard}, {Arnouts}, {Ilbert}, {Coupon},
  {Davidzon}, {Guzzo}, {Hudelot}, {McCracken}, {Van Werbaeke}, {Morrison}, {Le
  F{\`e}vre}, {Comte}, {Bolzonella}, {Fritz}, {Garilli}, \&
  {Scodeggio}}]{moutard16b}
{Moutard}, T., {Arnouts}, S., {Ilbert}, O., {et~al.} 2016b, \aap, 590, A103

\bibitem[{{Moutard} {et~al.}(2016a){Moutard}, {Arnouts}, {Ilbert}, {Coupon},
  {Hudelot}, {Vibert}, {Comte}, {Conseil}, {Davidzon}, {Guzzo}, {Llebaria},
  {Martin}, {McCracken}, {Milliard}, {Morrison}, {Schiminovich}, {Treyer}, \&
  {Van Werbaeke}}]{moutard16a}
{Moutard}, T., {Arnouts}, S., {Ilbert}, O., {et~al.} 2016a, \aap, 590, A102

\bibitem[{{Pezzotta} {et~al.}(2016)}]{pezzotta16}
{Pezzotta}, A. {et~al.} 2016, \aap~\rm{submitted}, \rm{ArXiv e-print XXXX.YYYY}

\bibitem[{{Puget} {et~al.}(2004){Puget}, {Stadler}, {Doyon}, {Gigan},
  {Thibault}, {Luppino}, {Barrick}, {Benedict}, {Forveille}, {Rambold},
  {Thomas}, {Vermeulen}, {Ward}, {Beuzit}, {Feautrier}, {Magnard}, {Mella},
  {Preis}, {Vallee}, {Wang}, {Lin}, {Hall}, \& {Hodapp}}]{Puget2004}
{Puget}, P., {Stadler}, E., {Doyon}, R., {et~al.} 2004, in \procspie, Vol.
  5492, Ground-based Instrumentation for Astronomy, ed. A.~F.~M. {Moorwood} \&
  M.~{Iye}, 978--987

\bibitem[{{Rota} {et~al.}(2016)}]{rota16}
{Rota}, S. {et~al.} 2016, \aap~\rm{in press}, \rm{arxiv:XXXX.YYYY}

\bibitem[{{Wilson} {et~al.}(2016)}]{wilson16}
{Wilson}, M. {et~al.} 2016, \aap~\rm{submitted}, \rm{arXiv:XXXX.YYYY}

\bibitem[{{York} {et~al.}(2000){York}, {Adelman}, {Anderson}, {Anderson},
  {Annis}, {Bahcall}, {Bakken}, {Barkhouser}, {Bastian}, {Berman}, {Boroski},
  {Bracker}, {Briegel}, {Briggs}, {Brinkmann}, {Brunner}, {Burles}, {Carey},
  {Carr}, {Castander}, {Chen}, {Colestock}, {Connolly}, {Crocker}, {Csabai},
  {Czarapata}, {Davis}, {Doi}, {Dombeck}, {Eisenstein}, {Ellman}, {Elms},
  {Evans}, {Fan}, {Federwitz}, {Fiscelli}, {Friedman}, {Frieman}, {Fukugita},
  {Gillespie}, {Gunn}, {Gurbani}, {de Haas}, {Haldeman}, {Harris}, {Hayes},
  {Heckman}, {Hennessy}, {Hindsley}, {Holm}, {Holmgren}, {Huang}, {Hull},
  {Husby}, {Ichikawa}, {Ichikawa}, {Ivezi{\'c}}, {Kent}, {Kim}, {Kinney},
  {Klaene}, {Kleinman}, {Kleinman}, {Knapp}, {Korienek}, {Kron}, {Kunszt},
  {Lamb}, {Lee}, {Leger}, {Limmongkol}, {Lindenmeyer}, {Long}, {Loomis},
  {Loveday}, {Lucinio}, {Lupton}, {MacKinnon}, {Mannery}, {Mantsch}, {Margon},
  {McGehee}, {McKay}, {Meiksin}, {Merelli}, {Monet}, {Munn}, {Narayanan},
  {Nash}, {Neilsen}, {Neswold}, {Newberg}, {Nichol}, {Nicinski}, {Nonino},
  {Okada}, {Okamura}, {Ostriker}, {Owen}, {Pauls}, {Peoples}, {Peterson},
  {Petravick}, {Pier}, {Pope}, {Pordes}, {Prosapio}, {Rechenmacher}, {Quinn},
  {Richards}, {Richmond}, {Rivetta}, {Rockosi}, {Ruthmansdorfer}, {Sandford},
  {Schlegel}, {Schneider}, {Sekiguchi}, {Sergey}, {Shimasaku}, {Siegmund},
  {Smee}, {Smith}, {Snedden}, {Stone}, {Stoughton}, {Strauss}, {Stubbs},
  {SubbaRao}, {Szalay}, {Szapudi}, {Szokoly}, {Thakar}, {Tremonti}, {Tucker},
  {Uomoto}, {Vanden Berk}, {Vogeley}, {Waddell}, {Wang}, {Watanabe},
  {Weinberg}, {Yanny}, \& {Yasuda}}]{york00}
{York}, D.~G., {Adelman}, J., {Anderson}, Jr., J.~E., {et~al.} 2000, \aj, 120,
  1579

\end{thebibliography}

\end{document}